\documentclass[journal]{IEEEtran}
\usepackage{graphicx}
\usepackage{amsmath}
\usepackage{amssymb}
\usepackage[caption = false]{subfig}
\usepackage[mathscr]{eucal}

\usepackage{multirow}
\usepackage{amsmath,amssymb,bm}
\usepackage{amsfonts,dsfont,color,bbm}
\usepackage{algorithm}
\usepackage{cite,epstopdf,epsf,epsfig}
\usepackage{algpseudocode}
\usepackage{pifont}
\usepackage{amsmath}
\usepackage{booktabs}
\usepackage[bottom]{footmisc}

\DeclareMathOperator{\sign}{sign}

\ifCLASSINFOpdf
\else
\fi
\begin{document}

\title{Multimodal Deep Unfolding for Guided Image Super-Resolution}

\author{Iman~Marivani,~\IEEEmembership{Student Member,~IEEE,}
        Evaggelia~Tsiligianni,~\IEEEmembership{Member,~IEEE,}
        Bruno~Cornelis,~\IEEEmembership{Member,~IEEE,}
        and~Nikos~Deligiannis,~\IEEEmembership{Member,~IEEE}% <-this % stops a space
\thanks{The authors  are  with  the  Department  of  Electronics  and  Informatics, Vrije Universiteit Brussel, 1050 Brussels and with imec, Kapeldreef 75, 3001 Leuven, Belgium. (e-mail: \{imarivan, etsiligi, bcorneli, ndeligia\}@etrovub.be)}% <-this % stops a space
}


\markboth{}%
{Shell \MakeLowercase{\textit{et al.}}: Bare Demo of IEEEtran.cls for IEEE Journals}



\maketitle

\begin{abstract}
The reconstruction of a high resolution image given a low resolution observation
is an ill-posed inverse problem in imaging.
Deep learning methods rely on training data
to learn an end-to-end mapping from a low-resolution input 
to a high-resolution output. 
Unlike existing deep multimodal models 
that do not incorporate domain knowledge about the problem,
we propose a multimodal deep learning design that
incorporates sparse priors and allows the effective integration of information
from another image modality into the network architecture.
Our solution relies on a novel deep unfolding operator,
performing steps similar to an iterative algorithm 
for convolutional sparse coding with side information;
therefore, the proposed neural network is interpretable by design.
The deep unfolding architecture is used as a core component of a multimodal framework for guided image super-resolution.
An alternative multimodal design is investigated
by employing residual learning to improve the training efficiency.
The presented multimodal approach is applied to super-resolution of near-infrared 
and multi-spectral images as well as depth upsampling using RGB\ images as side information.
Experimental results show that our model outperforms state-of-the-art methods.
\end{abstract}


% \fi
\IEEEpeerreviewmaketitle

\section{Introduction}
\label{sec:intro}
\IEEEPARstart{I}{mage} super-resolution (SR) is a well-known inverse problem in imaging, 
referring to the reconstruction of a high-resolution (HR) image from a low-resolution (LR) observation~\cite{ribes2008linear, lucas2018using}.
The problem is ill-posed as there is no unique mapping from the LR to the HR image.
Practical applications such as medical imaging and remote sensing
often involve different image modalities capturing the same scene,
therefore, another approach in imaging is the joint use of multiple image modalities.
The problem of multimodal or guided image super-resolution
refers to the reconstruction of an HR image from an LR observation 
using a \textit{guidance} image from another modality,
also referred to as \textit{side information}. 

Several image processing methods use prior knowledge about the image
such as sparse structure~\cite{sparse1, sparse3, sparse4, sparse5, yang2008image, jia2012image , CSCSR} 
or statistical image priors~\cite{kim2010single, interpol3}.
Deep learning has been widely used in inverse problems, 
often outperforming analytical methods~\cite{lucas2018using, nguyen2017deep}.
For instance, in single image SR,
Convolutional Neural Networks (CNNs) have led to impressive results 
reported in~\cite{dong2016image, dong2016accelerating, 15prim, kim2016deeply, mao2016image, Bruna2015}.
Residual learning~\cite{res} enabled the training of very deep neural networks (DNNs), 
with the models proposed in~\cite{EDSR, 51prim, DRRN, memNet, RDN} 
achieving state-of-the-art performance.

Capturing the correlation among different image modalities
has been addressed with sparsity-based analytical models
and coupled dictionary learning~\cite{wang2012semi, zhuang2013supervised,
liu2014semi, jing2015super, dao2016collaborative, bahrampour2015multimodal, deligiannis2016multi, deligiannis2016x, MSR}.
The main drawback of these approaches is the 
high-computational cost of iterative algorithms for sparse approximation,
which has been addressed by multimodal deep learning methods~\cite{ngiam2011multimodal, DJF, DGF}.
A common approach in multimodal neural network design
is the fusion of the input modalities at a shared latent layer, 
obtained as the concatenation of the latent representations of each modality~\cite{ngiam2011multimodal}.
The CNN model proposed in~\cite{DJF} for multimodal depth upsampling follows this principle. 
Nevertheless, current multimodal DNNs are black-box models in the sense that 
we lack a principled approach to design such models 
for leveraging the signal structure and properties of the correlation across modalities.

A recent line of research in deep learning for inverse problems
considers \textit{deep unfolding}~\cite{LISTA, hershey2014deep, xin2016maximal, borgerding2017amp, ADMM-Net, raja},
that is, the unfolding of an iterative algorithm into the form of a DNN.
Inspired by numerical algorithms for sparse coding, 
deep unfolding designs have been utilized in several imaging problems  
to incorporate sparse priors into the solution.
Results for denoising~\cite{raja}, compressive imaging~\cite{zhang2018ista}, 
and image SR~\cite{Huang}
have shown that incorporating domain knowledge  
into the network architecture can improve the performance substantially. 
Still, these methods focus on single-modal data, thereby lacking a principled way
to incorporate knowledge from different imaging modalities.
To the best of our knowledge, the only deep unfolding designs for guided image SR
have been presented in~\cite{JMDL} and~\cite{CUNet},
that build upon existing unfolding architectures for learned sparse coding~\cite{LISTA} and learned convolutional sparse coding~\cite{raja}, respectively.  

In this paper, we address the problem of guided image SR
with a novel multimodal deep unfolding architecture,
which is inspired by a proximal algorithm for convolutional sparse coding with side information. 
The proximal algorithm is translated into a neural network form 
coined Learned Multimodal Convolutional Sparse Coding (LMCSC);
the network incorporates sparse priors and 
enables efficient integration of the guidance modality into the solution.
While in existing multimodal deep learning methods~\cite{ngiam2011multimodal, DJF, DGF},
it is difficult to understand what the model has learned,
our deep neural network is interpretable,
in the sense that the model performs steps similar to an iterative algorithm.

The proposed approach builds upon our previous research on multimodal deep unfolding~\cite{lesitaSPL, lesitaEUSIPCO},
and preliminary results of LMCSC can be found in~\cite{lesitaICIP},
where the recovery of HR near infrared (NIR) images based on LR NIR observations with the aid of RGB images was addressed.
In this paper, we integrate LMCSC into different neural network architectures, 
and present experiments on various multimodal datasets, 
showing the superior performance of the proposed approach
against several single- and multimodal SR methods.
Our contribution is as follows:
\begin{description}
	\item[(i)] We formulate the problem of convolutional sparse coding with side information, 
	and propose a proximal algorithm for its solution.
	\item[(ii)] Inspired by the proposed proximal algorithm, we design a deep unfolding neural network 
	for fast computation of convolutional sparse codes with the aid of side information.
	\item[(iii)] The deep unfolding operator is used as a core component 
	in a novel multimodal framework for guided image SR
	that fuses information from two image modalities.
	Furthermore, we exploit residual learning and introduce skip connections in the proposed framework,
	obtaining an alternative design that can be trained more efficiently.
	\item[(iv)] We test our models on several benchmark multimodal datasets, 
	including NIR/RGB, multi-spectral/RGB, depth/RGB, 
	and  compare them against various state-of-the-art single-modal and multimodal models. The numerical results show a PSNR gain of up to $2.74$dB over the coupled ISTA method in~\cite{JMDL}.
\end{description}

%
%
%\begin{itemize}
%\item{We formulate the problem of convolutional sparse coding with side information, 
%and propose a proximal algorithm for its solution.}
%\item{Inspired by the proposed proximal algorithm, we design a deep unfolding neural network 
%for fast computation of convolutional sparse codes with the aid of side information.
%}
%\item{The deep unfolding operator is used as a core component 
%in a novel multimodal framework for image SR
%that fuses information from two image modalities.}
%\item{We exploit residual learning and introduce skip connections in the proposed framework,
%obtaining alternative designs that can be trained more efficiently.}
%\end{itemize}
%% 

The rest of the paper is organized as follows: 
Section~\ref{sec:related} reviews related work and Section~\ref{sec:back} provides the necessary background.
Section~\ref{sec:LMSC} presents the proposed core deep unfolding architecture for convolutional sparse coding with side information, and our designs for multimodal image SR are presented in Section~\ref{sec:SRnets}, followed by experimental results in Section~\ref{sec:experiment}.
Section~\ref{sec:conclusion} concludes the paper.

Throughout the paper, all vectors are denoted by boldface lower case letters while lower case letters are used for scalars. We utilize boldface upper case letters to show matrices and boldface upper case letters in math calligraphy to indicate tensors. Moreover, in this paper, the terms upscaling factor and scale are used interchangeably.

%%%%%%%%%%%%%%%%%%%%%%%%%%%%%%%%%%
\section{Related Work}
\label{sec:related}
%%%%%%%%%%%%%%%%%%%%%%%%%%%%%%%%%%
\subsubsection{Single Image Super-Resolution}
A first category of single image SR methods includes 
interpolation-based methods~\cite{interpol3, sanchez2008noniterative, zhou2012interpolation, ling2013interpolation}.
These methods are simple and fast, however, aliasing and blurring effects make them inefficient 
in obtaining HR images of fine quality.
A second category includes reconstruction methods~\cite{yang2013fast, dong2012nonlocally, mallat2010super},
which use several image priors to regularize the ill-posed reconstruction problem
and result in images with fine texture details.
Nevertheless, modelling the complex context of natural images with image priors is not an easy task.
A third popular category consists of learning-based 
methods~\cite{sparse1, sparse3, sparse4, sparse5, yang2008image,  jia2012image, dong2016image,
dong2016accelerating, 15prim, kim2016deeply, mao2016image, Bruna2015, EDSR, 51prim, DRRN, memNet, RDN},
which use machine learning techniques to learn the complex mapping between LR and HR images from data.

Among learning based methods, deep learning models have drawn considerable attention
as they achieve excellent restoration quality.
%CNNs
SRCNN~\cite{dong2016image} was the first deep learning method for image SR.
The model has a simple structure and can directly learn 
an end-to-end mapping between the LR/HR images. 
%The first layer accepts as input a bicubic interpolation of an LR image,
%the second layer learns a non-linear mapping from the LR to the HR space,
%and the last layer performs the reconstruction of the HR image.
An accelerated version of~\cite{dong2016image} was presented in~\cite{dong2016accelerating}.
Increasing the depth of CNN architectures ensues several training difficulties 
which have been mitigated with residual learning.
Examples of very deep residual networks for SR
include a $20$-layer CNN proposed in~\cite{15prim, kim2016deeply},  
a $30$-layer convolutional autoencoder proposed in~\cite{mao2016image},
and a $52$-layer CNN proposed in~\cite{DRRN}.
Residual learning has also been employed to learn inter-layer dependencies in~\cite{memNet, RDN}.
%The residual structure presented in~\cite{memNet} is  combined with a gating mechanism 
%to capture the long-term dependencies between network layers.
%Following~\cite{memNet}, the authors of~\cite{RDN}
%propose a local feature fusion with a local residual learning architecture
%to make use of all the hierarchical features. % from the original dense connected layers.
An improved residual design obtained 
by removing unnecessary residual modules was proposed in~\cite{EDSR}.  
%RNNs
Recurrent neural networks (RNNs) have been also used for image SR in~\cite{SRFBN, han2018image}.
The network in~\cite{SRFBN}  implements a feedback mechanism 
that carries high-level information back to previous layers, 
refining low-level encoded information.
Following~\cite{SRFBN}, the authors of~\cite{han2018image} 
presented an alternative structure with two states (RNN layers) 
that operate at different spatial resolutions,
providing information flow from LR to HR encodings.

Deep unfolding has been applied to single image SR in~\cite{Huang}
where the authors designed a neural network 
that computes latent representations of the LR/HR image using LISTA~\cite{LISTA},
a neural network that performs steps similar to 
the Iterative Soft Thresholding Algorithm (ISTA)~\cite{ISTA} (see also Section~\ref{sec:back}).

%MULTIMODAL METHODS
\subsubsection{Multimodal Image Super-Resolution}
A common approach in multimodal image restoration is the joint or guided filtering approach,
that is, the design of a filter that leverages the guidance image as a prior
and transfers structural details from the guidance to the target image.
Several joint filtering techniques have been proposed in~\cite{JBU, MSR2, SDF}.
Nevertheless, when the local structures in the guidance and target images are not consistent, 
these techniques may transfer incorrect content to the target image.
%Another drawback of this approach concerns the explicit construction of the filter;
The approach presented in~\cite{JFSM} concerns the design of an explicit mapping
that captures the structural discrepancy between images from different modalities.
%the mapping is optimized with respect to adaptive smoothing, edge preservation and guidance strength manipulation.

Model-based techniques and joint filtering methods
are limited in characterizing the complex dependency between different modalities.
Learning based methods aim to learn this dependency from data.
In a depth upsampling method presented in~\cite{dynamic-guidance}, 
a weighted analysis representation is used to model the complex relationship 
between depth and RGB images;
the model parameters are learned with a task driven training strategy.
%SPARSITY BASED METHODS
Another learning based approach 
relies on sparse modelling and involves coupled-dictionary learning~\cite{wang2012semi, zhuang2013supervised, liu2014semi, jing2015super, dao2016collaborative, bahrampour2015multimodal, deligiannis2016multi, deligiannis2016x}.
Most of these works assume that there is a mapping between the sparse representation
of one modality to the sparse representation of another modality.
The authors of~\cite{MSR} consider both similarities and disparities between different modalities 
under the sparse representation invariance assumption.

Purely data-driven solutions for multimodal image SR are provided by 
multimodal deep learning approaches.
Examples include the model presented in~\cite{DJF} % Deep Joint Filtering 
implementing a CNN based joint image filter, 
and the work presented in~\cite{DGF},
which is a deep learning reformulation of the widely used guided image filter proposed in~\cite{MSR2}.

The deep unfolding design LISTA~\cite{LISTA} 
has also been deployed for multimodal image SR in~\cite{JMDL}
where a coupled ISTA network is presented.
The network accepts as input an LR image from the target modality
and an HR image from the guidance modality.
Two LISTA branches are employed to compute latent representations of the input images. 
The estimation of the target HR image is obtained  
as a linear combination of these representations. A similar approach is proposed in~\cite{CUNet} that employs three convolutional LISTA networks to split the common information shared between modalities, from the unique information belonging to each modality. The output is then computed as a combination of these common and unique feature maps after applying the corresponding dictionaries on them.

%%%%%%%%%%%%%%%%%%%%%%%%%%
\section{Background}
\label{sec:back}
%%%%%%%%%%%%%%%%%%%%%%%%%%
%%%%%%%%%%%%%%%%%%%%%%%%%%

Image super-resolution can be addressed as a linear inverse problem
formulated as follows~\cite{sparse1}:
\begin{equation}
\label{eq:IP}
\boldsymbol{y} = \boldsymbol{L} \boldsymbol{x} + \boldsymbol{\eta},
\end{equation}
where $\boldsymbol{x} \in \mathbb{R}^k$ is a vectorized form of the unknown HR image,
and $\boldsymbol{y} \in \mathbb{R}^n$ denotes the LR observations 
contaminated with noise $\boldsymbol{\eta} \in \mathbb{R}^n$.
The linear operator $\boldsymbol{L} \in \mathbb{R}^{n \times k}$, $n < k$, 
describes  the observation mechanism,
which can be expressed as the product of 
a downsampling operator $\boldsymbol{E}$ and a blurring filter $\boldsymbol{H}$~\cite{sparse1}.
Problem~\eqref{eq:IP} appears in many imaging applications
including image restoration and inpainting~\cite{ribes2008linear, lucas2018using}.

\subsection{Image Super-Resolution via Sparse Approximation}

Even when the linear observation operator $\boldsymbol{L}$ is given, 
problem~\eqref{eq:IP} is ill-posed and requires additional regularization for its solution.
Sparsity has been widely used as a regularizer
leading to the well-known sparse approximation problem~\cite{tropp2010computational}. 
Instead of directly solving for $\boldsymbol{x}$, in this paper, 
we rely on a sparse modelling approach presented in~\cite{sparse1}.
According to~\cite{sparse1},
an $n$-dimensional (vectorized) patch $\boldsymbol{y}$ from a bicubic-upscaled LR image
and the corresponding patch $\boldsymbol{x}$ from the respective HR image
can be expressed by joint sparse representations.
By jointly learning two  dictionaries $\boldsymbol{\Psi}_y \in \mathbb{R}^{n \times m}$, $\boldsymbol{\Psi}_x \in \mathbb{R}^{n \times m}$, $n \leq m$,
for the low- and the high-resolution image patches, respectively,
we can enforce the similarity of sparse representations of patch pairs
such that $\boldsymbol{y} = \boldsymbol{\Psi}_y\boldsymbol{\alpha}$ and $\boldsymbol{x} = \boldsymbol{\Psi}_x\boldsymbol{\alpha}$, $\boldsymbol{\alpha} \in \mathbb{R}^{m}$.
Then, computing the HR patch $\boldsymbol{x}$ is equivalent to 
finding the sparse representation of the LR patch $\boldsymbol{y}$, 
by solving
\begin{equation}
\label{eq:SC}
\min_{\boldsymbol{\alpha}}  \frac{1}{2}\| \boldsymbol{y} - \boldsymbol{\Psi}_y \boldsymbol{\alpha}\|^2_2+\lambda \|\boldsymbol{\alpha}\|_1,
\end{equation}
where $\lambda$ is a regularization parameter, 
and $\|\boldsymbol{\alpha}\|_1 = \sum_{i=1}^{k} |\alpha_i|$ is the $\ell_1$-norm, 
which promotes sparsity.
Several methods have been proposed for solving~\eqref{eq:SC} including pivoting algorithms, interior-point methods and  gradient based methods~\cite{tropp2010computational}.

%%%%%%%%%%%%%%%%%%%%%%%%%%%
%\subsection{Convolutional Sparse Coding}
%%%%%%%%%%%%%%%%%%%%%%%%%%%
Sparse modelling techniques that involve computations applied to independent image patches
do not take into account the consistency of pixels in overlapping patches~\cite{CSCSR}.
Convolutional Sparse Coding (CSC)~\cite{CSC} is an alternative approach,
which can be directly applied to the entire image. 
Denoting with $\boldsymbol{Y} \in \mathbb{R}^{n_1 \times n_2}$ the image of interest,
the convolutional sparse codes are obtained by solving the following problem:
\begin{equation}
\label{eq:CSC}
\min_{\boldsymbol{U}_i} \frac{1}{2}\| \boldsymbol{Y} - \sum\limits_{i=1}^{k}\boldsymbol{D}_i*\boldsymbol{U}_i\|^2_F+\lambda\sum\limits_{i=1}^{k}\|\boldsymbol{U}_i\|_1,
\end{equation} 
where $\|\boldsymbol{A}\|_F=\sqrt{\sum_{i}\sum_{j}|a_{ij}|^2}$ is the Frobenius norm, $\boldsymbol{D}_i \in \mathbb{R}^{p_1 \times p_2}$, $i=1,...,k$,  are the atoms
of a convolutional dictionary $\boldsymbol{\mathcal{D}} \in \mathbb{R}^{p_1 \times p_2 \times k}$,
and $\boldsymbol{U}_i \in \mathbb{R}^{n_1 \times n_2}$, $i=1,...,k$, are the sparse feature maps with respect to $\boldsymbol{\mathcal{D}}$. 
The $\ell_1$-norm computes the sum of absolute values of the elements 
in  $\boldsymbol{U}_i$ (as if $\boldsymbol{U}_i$ is unrolled as a vector). 
Efficient solutions of~\eqref{eq:CSC} are presented in~\cite{CSCADMM, heide2015fast}.

%%%%%%%%%%%%%%%%%%%%%%%%%%%
%\subsection{Sparse Approximation with Side Information}
%\label{sec:l1l1}
%%%%%%%%%%%%%%%%%%%%%%%%%%%
According to recent studies~\cite{nikosIT}, 
the accuracy of sparse approximation problems can be improved 
if a signal $\boldsymbol{\omega}$  correlated with the target signal $\boldsymbol{y}$ is available;
we refer to $\boldsymbol{\omega}$ as side information (SI).
Assume that $\boldsymbol{y} \in \mathbb{R}^{n}$ and $\boldsymbol{\omega} \in  \mathbb{R}^{d}$ 
have similar sparse representations $\boldsymbol{\alpha} \in \mathbb{R}^{m}$, $\boldsymbol{s} \in \mathbb{R}^{m}$, 
under dictionaries $\boldsymbol{\Psi}_y \in \mathbb{R}^{n \times m}$, $\boldsymbol{\Psi}_\omega \in \mathbb{R}^{d \times m}$, $n \leq m$, $d \leq m$, respectively.
Then the sparse representation $\boldsymbol{\alpha}$ can be obtained 
as the solution of the $\ell_1$-$\ell_1$ minimization problem~\cite{nikosIT}.
\begin{equation}
\min_{\boldsymbol{\alpha}} \frac{1}{2}\| \boldsymbol{y} - \boldsymbol{\Psi}_y \boldsymbol{\alpha}\|_2^2  + \lambda (\|\boldsymbol{\alpha}\|_1  +  \|\boldsymbol{\alpha}-\boldsymbol{s}\|_1).
\label{eq:l1l1}
\end{equation}
Problem~\eqref{eq:l1l1} has been theoretically studied in~\cite{nikosIT}.
Numerical methods for its solution are presented in~\cite{weizman2015compressed, mota2015dynamic}.

%%%%%%%%%%%%%%%%%%%%%%%%%%
\subsection{Deep Unfolding}
\label{sec:unfolding_back}
%%%%%%%%%%%%%%%%%%%%%%%%%%
Analytical approaches for sparse approximation are usually equipped with theoretical guarantees;
however, their major drawback is their high computational complexity.  
In some applications, the deployed dictionaries also need to be learned,
increasing the computational burden~\cite{sparse1}.
The authors of~\cite{LISTA} address this problem
by a neural network design that performs operations
similar to the Iterative Soft Thresholding Algorithm (ISTA)~\cite{ISTA}
proposed for the solution of~\eqref{eq:SC}.
The learning process results in a trained version of ISTA, coined LISTA.
The $t$-th layer of LISTA computes:
\begin{equation}
\label{eq:LISTA} 
\boldsymbol{\alpha}^{t} = \phi_{\gamma}(\boldsymbol{S}\boldsymbol{\alpha}^{t-1} + \boldsymbol{W}\boldsymbol{y}) , \quad \boldsymbol{\alpha}^{0}=0,
\end{equation}
where 
\begin{equation}
\label{eq:LISTAprox}
\phi_{\gamma}(v_i) = \text{sign}(v_i)\max\{0, |v_i| - \gamma\}, \quad  i=1, \dots , k, 
\end{equation}
is the soft thresholding operator;
$\boldsymbol{S}\in\mathbb{R}^{m \times m}$, $\boldsymbol{W}\in\mathbb{R}^{m \times n}$ and $\gamma>0$ are parameters,
which are fixed in ISTA, while LISTA learns them from data.
As a result, LISTA achieves high accuracy in only a few iterations.
The technique known as deep unfolding 
was also explored in~\cite{hershey2014deep, xin2016maximal, borgerding2017amp, ADMM-Net};
a convolutional LISTA design for CSC was presented in~\cite{raja}.

The aforementioned deep unfolding studies deal with single-modal data. 
A  deep unfolding design that incorporates side information coming from another modality
was first presented in our previous work~\cite{lesitaSPL}.
The model proposed in~\cite{lesitaSPL} relies on
a proximal method for the solution of \eqref{eq:l1l1},
which iterates over
\begin{equation}
\label{eq:sita}
\boldsymbol{\alpha}^{t} = \xi_{\mu}\big((\boldsymbol{I}- \frac{1}{L}\boldsymbol{\Psi}^T\boldsymbol{\Psi})\boldsymbol{\alpha}^{t-1} + \frac{1}{L}\boldsymbol{\Psi}^T\boldsymbol{y};\boldsymbol{s}\big), \quad \boldsymbol{\alpha}^{0}=0,
\end{equation}
with $\mu$, $L$ appropriate parameters. 
The proximal operator  $\xi_\mu$ 
incorporates the side information $\boldsymbol{s}$ and is expressed as follows:
\begin{enumerate}
\small
\item{For $s_i \ge 0$, $i=1, \dots, m$:
\begin{equation}
\small
\label{eq:prox1}
\xi_{\mu}(v_i;s_i) =  
\begin{cases}
\small
v_i + 2\mu, \, \; \; \, \; \; \qquad  v_i < -2\mu  \\
0, \quad \qquad \; \; \, \; \;  -2\mu \le v_i \le 0\\
v_i , \qquad \qquad \, \; \; \;  0 < v_i < s_i \\  
s_i , \qquad \;  \, s_i \le v_i \le s_i +2\mu \\
v_i - 2\mu , \, \qquad  v_i \ge s_i +2\mu
\end{cases}
\end{equation}
}
\item{For $s_i < 0$, $i=1, \dots, m$:
\begin{equation}
\small
\label{eq:prox2}
\xi_{\mu}(v_i;s_i) =  
\begin{cases}
\small
v_i + 2\mu,  \qquad  v_i < s_i -2\mu  \\
s_i , \qquad \;   s_i - 2\mu \le v_i \le s_i \\
v_i , \qquad \; \; \; \qquad  s_i < v_i < 0 \\ 
0, \qquad \quad \qquad  0 \le v_i \le 2\mu \\
v_i - 2\mu , \, \qquad \qquad  v_i \ge 2\mu  
\end{cases}.
\end{equation}
}
\end{enumerate}
By writing the proximal algorithm in the form
\begin{equation}
\label{eq:lesitaaa}
\boldsymbol{\alpha}^{t} = \xi_\mu(\boldsymbol{S}\boldsymbol{\alpha}^{t-1} + \boldsymbol{W}\boldsymbol{y};\boldsymbol{s}), \quad \boldsymbol{\alpha}^{0}=0,
\end{equation}
and translating~\eqref{eq:lesitaaa} into a deep neural network,
we obtain a fast multimodal operator referred to  
as Learned Side-Information-driven iterative soft Thresholding Algorithm (LeSITA).
%The parameterized by  $S$ and $W$ and $\mu$.
LeSITA has a similar expression to LISTA~\eqref{eq:LISTA}, 
however, \eqref{eq:lesitaaa} employs the new activation function~$\xi_\mu$ 
that integrates side information into the learning process. 

%%%%%%%%%%%%%%%%%%%%%%%%%%%%%%%%%%%%%%
\section{Design Multimodal Convolutional Networks with Deep Unfolding}
\label{sec:LMSC}
%%%%%%%%%%%%%%%%%%%%%%%%%%%%%%%%%%%%%%
\begin{figure}[t!]
	\centering
	\includegraphics[scale=0.26]{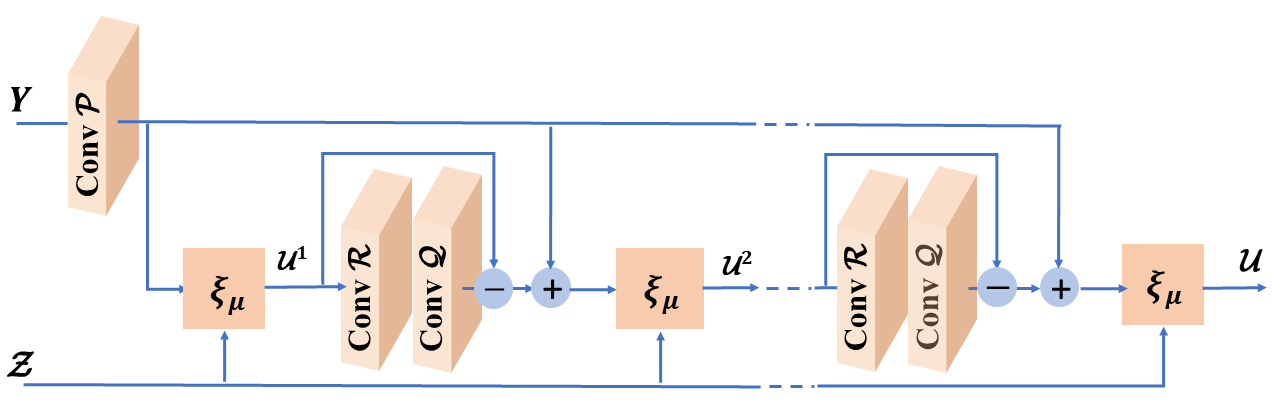}\\
	\caption{The proposed LMCSC model with unfolded recurrent stages. 
		The model computes sparse feature maps $\boldsymbol{\mathcal{U}}$ of an image $\boldsymbol{Y}$ given
		sparse feature maps $\boldsymbol{\mathcal{Z}}$ of the side information. 
		The nonlinear activation function follows the proximal operator $\xi_\mu(v;z)$ given by \eqref{eq:prox1},
		\eqref{eq:prox2}.}
	\label{fig:c_lesita}
\end{figure}
In what follows, we consider that,
besides the observations $\boldsymbol{Y}$ of the target signal, 
another image modality $\boldsymbol{\Omega}$, correlated with $\boldsymbol{Y}$ is available.
We assume that the two image modalities
can be represented by convolutional sparse codes
that are similar by means of the $\ell_1$-norm.
Specifically, let $\boldsymbol{Y} = \sum_{i=1}^{k}\boldsymbol{D}^Y_i*\boldsymbol{U}_i$
be a sparse representation of the observed image $\boldsymbol{Y}$
with respect to a convolutional dictionary $\boldsymbol{\mathcal{D}}^Y$;
$\boldsymbol{D}^{Y}_i \in \mathbb{R}^{n_1 \times n_2}$, $i=1,...,k$, denote the atoms of $\boldsymbol{\mathcal{D}}^Y$.
By employing a convolutional dictionary $\boldsymbol{\mathcal{D}}^\Omega$ with atoms $\boldsymbol{D}^{\Omega}_i \in \mathbb{R}^{p_1 \times p_2}$, $i=1,...,k$,
the guidance image  $\boldsymbol{\Omega} \in \mathbb{R}^{n_1 \times n_2}$
can be expressed as $\boldsymbol{\Omega} = \sum_{i=1}^{k}\boldsymbol{D}^{\Omega}_i*\boldsymbol{Z}_i$
with the convolutional sparse codes $\boldsymbol{Z}_i$, $i=1,...,k$, obtained as the solution of~\eqref{eq:CSC}.
Then, we can compute the unknown sparse codes $\boldsymbol{U}_i$ of the target modality
by solving a problem formulated in a way similar to~\eqref{eq:l1l1}, that is,
\begin{equation}
\label{eq:MCSC}
\min_{\boldsymbol{U}_i}  \frac{1}{2}\| \boldsymbol{Y} - \sum\limits_{i=1}^{k}\boldsymbol{D}^Y_i*\boldsymbol{U}_i\|^2_F+
\lambda(\sum\limits_{i=1}^{k}\|\boldsymbol{U}_i\|_1 +\sum\limits_{i=1}^{k}\|\boldsymbol{U}_i - \boldsymbol{Z}_i\|_1).
\end{equation}

There is a correspondence between convolutional and linear sparse codes.
If we replace the convolutional dictionary $\boldsymbol{\mathcal{D}}^Y$ with a matrix $\boldsymbol{A}$ with Toeplitz structure,
and take into account the linear properties of convolution,
then~\eqref{eq:MCSC} reduces to~\eqref{eq:l1l1}.
Specifically, we define $\boldsymbol{A} \in \mathbb{R}^{(n_1-p_1+1)(n_2-p_2+1) \times kn_1n_2}$ 
as a sparse dictionary obtained by concatenating the Toeplitz matrices that unroll $\boldsymbol{D}^Y_i$'s;
$\boldsymbol{\alpha} \in \mathbb{R}^{kn_1n_2}$ and $\boldsymbol{s} \in \mathbb{R}^{kn_1n_2}$ take the form of vectorized sparse feature maps 
of the target and the side information images, respectively.
Then, by replacing the convolutional operations in~\eqref{eq:MCSC} with multiplications, we obtain~\eqref{eq:l1l1},
and the proximal algorithm~\eqref{eq:sita} can be employed to compute convolutional sparse codes. 

Nevertheless, transforming~\eqref{eq:MCSC} to~\eqref{eq:l1l1} 
and using~\eqref{eq:sita} for its solution is not computationally efficient.  
Since CSC  deals with the entire image, the dimensionality of~\eqref{eq:l1l1} becomes too high
and the proximal method becomes impractical.
We use the correspondence between linear and convolutional representations,
and formulate an iterative algorithm that performs convolutions as follows:
In the proximal algorithm~\eqref{eq:sita}, the matrices $\boldsymbol{A}$ and $\boldsymbol{A}^T$, 
which take the form of concatenated Toeplitz matrices in the convolutional case,
are replaced by the convolutional dictionaries $\boldsymbol{\mathcal{B}}^Y$ and $\boldsymbol{\mathcal{\tilde{B}}}^Y$, respectively.
Then, by replacing multiplications with convolutional operations,
we can compute the convolutional codes $\boldsymbol{\mathcal{U}}$ of the target image, 
given the convolutional codes $\boldsymbol{\mathcal{Z}}$ of the guidance image,
by iterating over:
\begin{equation}
\label{eq:convsita}
\boldsymbol{\mathcal{U}}^{t} = \xi_{\mu}(\boldsymbol{\mathcal{U}}^{t-1} - \boldsymbol{\mathcal{\tilde{B}}}^Y*\boldsymbol{\mathcal{B}}^Y*\boldsymbol{\mathcal{U}}^{t-1}+\boldsymbol{\mathcal{\tilde{B}}}^Y*\boldsymbol{Y};\boldsymbol{\mathcal{Z}}),
\end{equation}
where $\boldsymbol{\mathcal{U}}$, $\boldsymbol{\mathcal{Z}}$ are tensors of size $p_1 \times p_2 \times k$.

Equation~\eqref{eq:convsita} can be translated into a deep convolutional neural network (CNN).
Each stage of the network computes the sparse feature maps according to  
\begin{equation}
\label{eq:lesita3}
\boldsymbol{\mathcal{U}}^{t} = \xi_{\mu}(\boldsymbol{\mathcal{U}}^{t-1} - \boldsymbol{\mathcal{Q}}*\boldsymbol{\mathcal{R}}*\boldsymbol{\mathcal{U}}^{t-1}+\boldsymbol{\mathcal{P}}*\boldsymbol{Y};\boldsymbol{\mathcal{Z}}),
\end{equation}
with $\boldsymbol{\mathcal{Q}} \in \mathbb{R}^{p_1 \times p_2 \times c \times k}$, 
$\boldsymbol{\mathcal{R}}  \in \mathbb{R}^{p_1 \times p_2 \times k \times c}$, 
$\boldsymbol{\mathcal{P}}  \in \mathbb{R}^{p_1 \times p_2 \times c \times k }$ learnable convolutional layers 
and $\mu>0$ a learnable parameter;
$c$ is the number of channels of the employed images.
The proposed network architecture, depicted in Fig.~\ref{fig:c_lesita}, 
is referred to as Learned Multimodal Convolutional Sparse Coding (LMCSC). 
The network can be trained in a supervised manner to map an input image to sparse feature maps.
During training, the parameters $\boldsymbol{\mathcal{Q}}$, $\boldsymbol{\mathcal{R}}$, $\boldsymbol{\mathcal{P}}$ and $\mu$ are learned;
therefore, the deep LMCSC can achieve high accuracy with only a fraction of computations of the proximal method.

\begin{figure}[t!]
	\centering
	\includegraphics[width=0.48\textwidth]{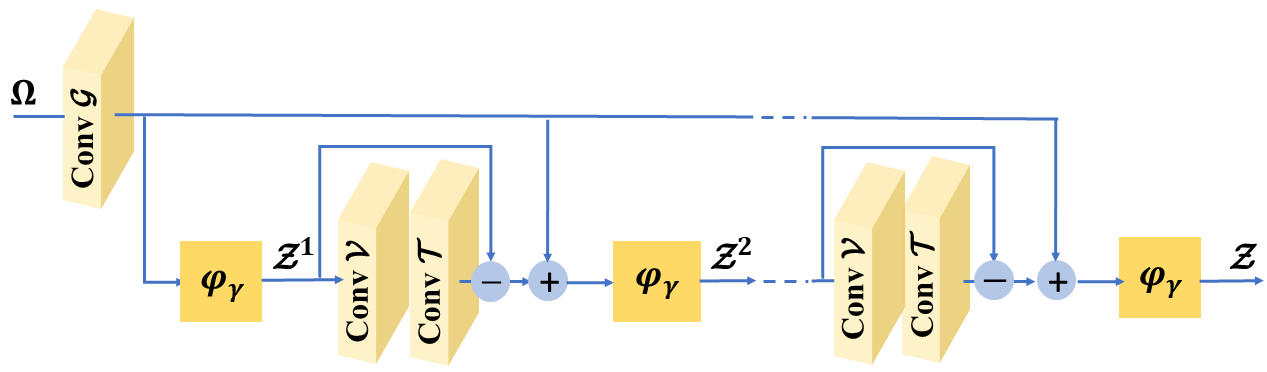}\\
	\caption{The Approximate Convolutional Sparse Coding (ACSC) model~\cite{raja} is used to encode the side information.
		The nonlinear activation function follows the proximal operator $ \phi_{\gamma}(v)$ in~\eqref{eq:LISTAprox}.}
	\label{fig:c_lista}
\end{figure}
LMCSC uses the convolutional sparse codes $\boldsymbol{\mathcal{Z}}$ of the guidance modality $\boldsymbol{\Omega}$
to compute the convolutional sparse codes of the target modality $\boldsymbol{Y}$.
An efficient multimodal convolutional operator should integrate a fast operator 
for the encoding of the guidance modality.
In the models presented next, 
we obtain $\boldsymbol{\mathcal{Z}}$ using the ACSC operator presented in~\cite{raja}.
ACSC has the form of convolutional LISTA, with the $t$-th layer computing:
\begin{equation}
\label{eq:ACSC}
\boldsymbol{\mathcal{Z}}^{t} = \phi_{\gamma}(\boldsymbol{\mathcal{Z}}^{t-1} - \boldsymbol{\mathcal{T}}*\boldsymbol{\mathcal{V}}* \boldsymbol{\mathcal{Z}}^{t-1} +\boldsymbol{\mathcal{G}}*\boldsymbol{\Omega} ),
\end{equation}
where $\phi_{\gamma}$ is the proximal operator given by~\eqref{eq:LISTAprox}.
The parameters of the convolutional layers $\boldsymbol{\mathcal{T}} \in \mathbb{R}^{p_1 \times p_2  \times c \times k}$, $\boldsymbol{\mathcal{G}} \in \mathbb{R}^{p_1 \times p_2  \times c \times k}$ and $\boldsymbol{\mathcal{V}} \in \mathbb{R}^{p_1 \times p_2  \times k \times c}$, are learned from data.
The architecture of ACSC is depicted in Fig.~\ref{fig:c_lista}.

%%%%%%%%%%%%%%%%%%%%%%%%%%%%%%%%%%
\section{Deep Multimodal Image Super-Resolution}
\label{sec:SRnets}
%%%%%%%%%%%%%%%%%%%%%%%%%%%%%%%%%%
The proposed LMCSC architecture can be employed 
to perform multimodal image super-resolution
based on a sparsity-driven convolutional model.
The proposed model follows similar principles with~\cite{sparse1}. 
Specifically, the sparse linear modelling of LR/HR image patches presented in~\cite{sparse1}
is replaced by a sparse convolutional modelling of the entire LR/HR images,
followed by similarity assumptions between the convolutional representations of the LR and HR images.
To efficiently integrate information from a second image modality,
we also assume that the target and the guidance image modalities
are similar by means of the $\ell_1$-norm in the representation domain.

%%
%Moreover, if the target and the guidance image modalities are highly correlated,
%we can assume that the convolutional sparse codes $z_i$
%of the guidance HR image $\Omega$ are similar to $\alpha_i$, for example, by means of the $\ell_1$ norm. 
%Then, the guidance image can aid the computation of $\alpha_i$
%via the formulation of the $\ell_1$-$\ell_1$ minimization problem~\eqref{eq:MCSC},
%with $d_i:=d^Y_i$, $i=1, \dots, k$.
%
%Clearly, a deep learning solution for multimodal image super-resolution
%that follows the presented convolutional modelling
%involves a neural network design that captures the correlation
%between the  target and the guidance image modalities.
%LMCSC can be used as a core component of a deep architecture
%that enables learning similar convolutional representations of correlated signals.
%Next, we present different variants of the proposed design and compare their performance.

%%%%%%%%%%%%%%%%%%%%%%%%%%%%%%%%%%
\subsection{LMCSC-Net}
\label{sec:model1}
%%%%%%%%%%%%%%%%%%%%%%%%%%%%%%%%%%
Our first model proposed for multimodal image super-resolution relies on the following assumption:
The LR observation $\boldsymbol{Y}$ and the HR image $\boldsymbol{X}$  
share the same convolutional sparse features maps  $\boldsymbol{U}_i$,
under different convolutional dictionaries $\boldsymbol{\mathcal{D}}^Y$ and $\boldsymbol{\mathcal{D}}^X$, 
that is, $\boldsymbol{Y} = \sum_{i=1}^{k}\boldsymbol{D}^Y_i*\boldsymbol{U}_i$, $\boldsymbol{X} = \sum_{i=1}^{k}\boldsymbol{D}^X_i*\boldsymbol{U}_i$,
where $\boldsymbol{D}^Y_i$, $\boldsymbol{D}^X_i$ are the atoms of the respective convolutional dictionaries.
Given $\boldsymbol{\mathcal{D}}^Y$, $\boldsymbol{\mathcal{D}}^X$, finding a mapping from $\boldsymbol{Y}$ to $\boldsymbol{X}$ 
is equivalent to computing the sparse features maps $\boldsymbol{\mathcal{U}}$ 
of the observed LR image $\boldsymbol{Y}$. 
The similarity assumption between the target and the guidance image modalities
in the representation domain implies that
the convolutional sparse codes $\boldsymbol{\mathcal{Z}}$ of the guidance HR image $\boldsymbol{\Omega}$ 
are similar to $\boldsymbol{\mathcal{U}}$ by means of the $\ell_1$-norm. 
Therefore, $\boldsymbol{\mathcal{U}}$ can be obtained as the solution of 
the $\ell_1$-$\ell_1$ minimization problem~\eqref{eq:MCSC}.
Based on these assumptions, we build our first model for multimodal image SR 
using LMCSC as a core component of a deep architecture.

The proposed model, coined LMCSC-Net, consists of three subnetworks:
(\textit{i}) an LMCSC encoder that produces convolutional latent representations of the imput LR image with the aid of side information,
(\textit{ii}) a side information encoder that produces latent representations of the guidance HR image,
and (\textit{iii}) a convolutional decoder that computes the target HR image. 
The goal of the LMCSC encoder is to learn a convolutional sparse feature map $\boldsymbol{\mathcal{U}}$
of the LR image $\boldsymbol{Y}$, also shared by the HR image $\boldsymbol{X}$,
using a convolutional sparse feature map $\boldsymbol{\mathcal{Z}}$ as side information, 
akin to the model presented in Section~\ref{sec:LMSC}. 
The LMCSC branch is followed by a convolutional decoder
realized by a learnable convolutional dictionary $\boldsymbol{\mathcal{D}}^X$.
The decoder receives the latent representations $\boldsymbol{\mathcal{U}}$ provided by LMCSC
and estimates $\boldsymbol{X}$ according to $\boldsymbol{\hat{X}} = \boldsymbol{\mathcal{D}}^X * \boldsymbol{\mathcal{U}}$.

%Since LMCSC accepts the side information in the form of convolutional sparse codes,
%a plausible question is how can we compute these codes efficiently?
%We employ the Approximate Convolutional Sparse Coding (ACSC) model presented in~\cite{raja} 
%to encode the side information.
%The model is a convolutional version of LISTA (see Fig.~\ref{fig:c_lista});
%therefore, it can be used to compute convolutional sparse codes of the guidance image $\Omega$ according to: 
%%
%\begin{equation}
%\label{eq:ACSC}
%z^{t+1} = \phi_{\gamma}(z^t - V*U* z^t +T*\Omega )),
%\end{equation}
%%
%where $\phi_{\gamma}$ is the proximal operator given by~\eqref{eq:LISTAprox};
%$T \in \mathbb{R}^{p_1 \times p_2  \times c \times k}$, $V \in \mathbb{R}^{p_1 \times p_2  \times c \times k}$, $U \in \mathbb{R}^{p_1 \times p_2  \times k \times c}$
%and $\gamma>0$ are learnable variables;
%$c$ is the number of channels of the employed images.

\begin{figure}[t!]
	\centering
	\includegraphics[width=0.45\textwidth]{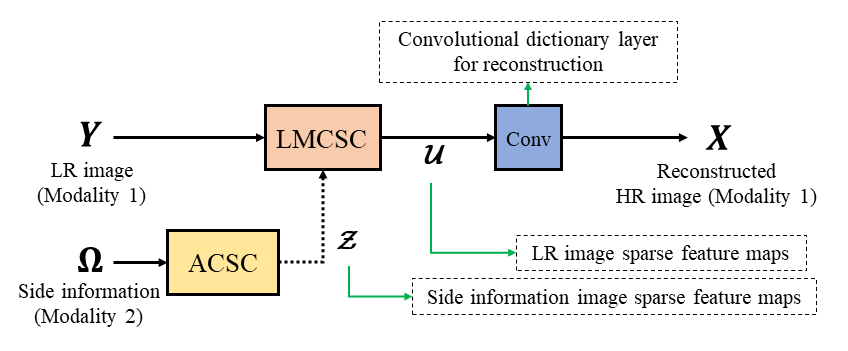}\\
	\caption{The proposed LMCSC-Net, a deep multimodal SR network consisting of an LMCSC encoder, an ACSC side information encoder, 
		and a convolutional decoder. }
	\label{fig:MCSC}
\end{figure}

The entire network, depicted in Fig.~\ref{fig:MCSC}, is trained end-to-end using 
the mean square error (MSE) loss function:
\begin{equation}
\label{eq:obj}
\min_{\Theta} \sum\limits_{i} \| \boldsymbol{\hat{X}}_i-\boldsymbol{X}_i\|_F^2,
\end{equation}   
where $\Theta$ denotes the set of all network parameters, 
$\boldsymbol{X}_i$ is the ground-truth image of the target modality,
and $\boldsymbol{\hat{X}}_i$ is the estimation computed by the network.

Different from our previous multimodal image SR design~\cite{lesitaEUSIPCO} 
which relies on LeSITA~\cite{lesitaSPL},
LMCSC-Net has a novel convolutional structure inspired by a different proximal algorithm.
The core LMCSC component computes latent representations of the target modality 
using side information from the guidance modality,
performing fusion of information at every layer.
Therefore, our approach is different from coupled ISTA~\cite{JMDL} 
which employs one branch of LISTA~\cite{LISTA} for each modality 
and fuses the latent representations only in the last layer.

%%%%%%%%%%%%%%%%%%%%%%%%%%%%%%%%%%
\subsection{LMCSC$_+$-Net}
%%%%%%%%%%%%%%%%%%%%%%%%%%%%%%%%%%
%
\begin{figure}[t!]
	\centering
	\includegraphics[width=0.47\textwidth]{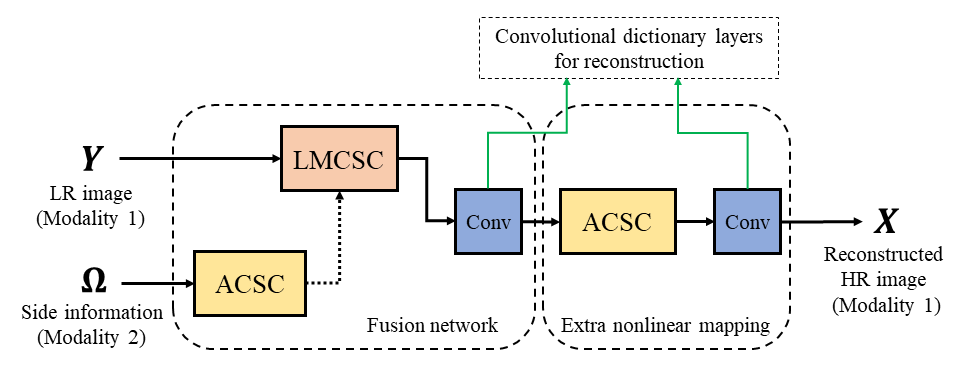}\\
	\caption{The proposed LMCSC$_+$-Net, a deep multimodal SR network consisting of an LMCSC component 
		with an additional ACSC branch performing enhancement of the LR/HR mapping. }
	\label{fig:LMCSC-SR+}
\end{figure}

The model presented in Section~\ref{sec:model1}
learns similar sparse representations of three different image modalities,
that is, the input LR image $\boldsymbol{Y}$, the guidance modality $\boldsymbol{\Omega}$ and the HR image $\boldsymbol{X}$.
Learning representations that mainly encode  the common information 
among the different modalities is critical for the performance of the model.
Nevertheless, some information from the guidance modality may be misleading
when learning a mapping between $\boldsymbol{X}$ and $\boldsymbol{Y}$. 
In other words, the encoding performed by the ACSC branch may result in transferring
unrelated information to the LMCSC encoder.
As a result, the latent representation of the target modality may not capture the  
underlying mapping between the LR and HR images in the representation domain. 
Furthermore, assuming identical latent representations for both LR and HR images 
limits the performance of the network 
especially when the degradation level of the LR observations is high. 

In order to address the aforementioned problems, 
we relax the assumption concerning the similarity between the LR and HR images in the representation domain.
Specifically,  we assume that the convolutional sparse codes $\boldsymbol{\mathcal{U}}_x$ of the HR image $\boldsymbol{X}$
can be obtained as a non-linear transformation of the respective codes $\boldsymbol{\mathcal{U}}_y$ of the LR image $\boldsymbol{Y}$,
that is, $\boldsymbol{\mathcal{U}}_x = F_{\Theta'}(\boldsymbol{\mathcal{U}}_y)$ where $F_{\Theta'}$ is a non-linear function parameterized by ${\Theta'}$. 

Under this assumption, we build the proposed multimodal SR framework
by employing the following components:
(i)~An LMCSC subnetwork is used to fuse the information 
from the LR observations and the guidance HR  image,
providing a first estimation of the target HR image with the aid of side information.
(ii)~An ACSC subnetwork following the LMCSC subnetwork
is used to enhance the transformation between the LR and HR images of the target modality
without using side information.
The architecture of the proposed model, referred to as LMCSC$_+$-Net, is depicted in Fig.~\ref{fig:LMCSC-SR+}. 
The additional ACSC and dictionary layers in LMCSC$_+$-Net implement $F_{\Theta'}$.
The network is trained using the objective~\eqref{eq:obj}.

%%%%%%%%%%%%%%%%%%%%%%%%%%%%%%%%%%
\subsection{LMCSC-Net with Skip Connections}
\begin{figure}
	\centering
	{\includegraphics[scale=0.32]{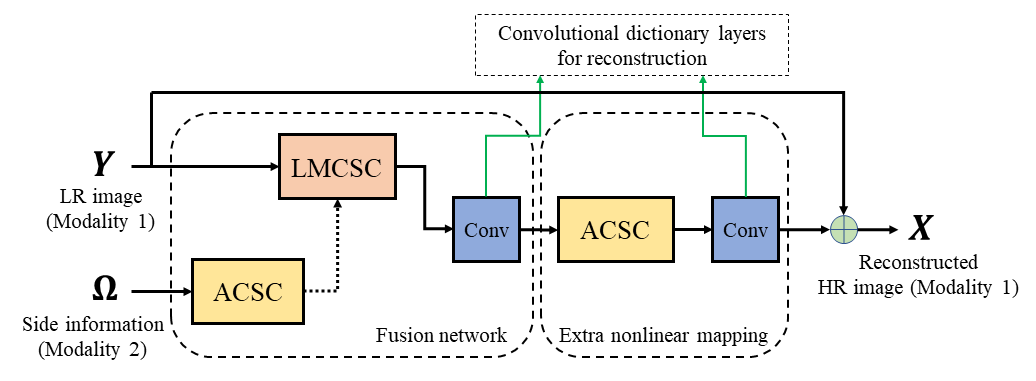}}
	\caption{The proposed LMCSC-ResNet, a deep multimodal LMCSC-based network with a skip connection. }   
	\label{fig:LMSCS_ResNets}
\end{figure}

%%%%%%%%%%%%%%%%%%%%%%%%%%%%%%%%%%
Considering the significant improvement achieved by residual learning 
in the training efficiency and the prediction accuracy~\cite{res,EDSR, 51prim, DRRN, memNet, RDN},
we enhance the proposed LMCSC$_+$-Net with a skip connection,
introducing a new model coined LMCSC-ResNet. 
For the design of LMCSC-ResNet we rely on the assumption that the HR image $\boldsymbol{X}$
contains all the low-frequency information from the LR image $\boldsymbol{Y}$ 
plus some high-frequency details that can be captured 
by a non-linear mapping between LR and HR images.
By using an identity mapping of the input, 
the capacity of the network can be assigned to learning the high frequency details,
since the low-frequency information is provided by $\boldsymbol{Y}$.
LMCSC-ResNet learns the non-linear mapping $H(\boldsymbol{Y}, \boldsymbol{\Omega}) = F(\boldsymbol{Y}, \boldsymbol{\Omega}) + \boldsymbol{Y}$,
where $F(\boldsymbol{Y}, \boldsymbol{\Omega}) = \boldsymbol{\hat{X}} - \boldsymbol{Y}$.
In LMCSC-ResNet, $F(\boldsymbol{Y}, \boldsymbol{\Omega})$ is obtained from the LMCSC$_{+}$-Net.
The model architecture is presented in Fig.~\ref{fig:LMSCS_ResNets}.
This model is also trained end-to-end using the objective~\eqref{eq:obj}.

%%%%%%%%%%%%%%%%%%%%%%%%%%%%%%%%%%
\section{Experiments}
\label{sec:experiment}
%%%%%%%%%%%%%%%%%%%%%%%%%%%%%%%%%%
%\begin{table}[t!]
%	\centering
%	\caption{Performance comparison [in terms of average PSNR (dB) over all test images] of different variants     of the proposed LMCSC-based framework for multimodal image SR 
%		for $\times 4$, $\times 6$ upscaling factors.}
%	\label{tab:table1}
%	\begin{center}
%		\addtolength{\tabcolsep}{4pt}
%		%	\addtolength{\tabcolsep}{-3pt}
%		\begin{tabular}{c || c | c | c }
%			\hline
%			Scale&{LMCSC-Net}&{LMCSC$_+$-Net}&{LMCSC-ResNet}\\
%			\hline
%			\hline
%			$\times4$ & $33.96$		& $34.23$		&$\bf34.36$\\
%			\hline
%			$\times6$ & $32.07$		&$31.94$ 		&$\bf32.33$\\
%			\hline
%			
%		\end{tabular}
%	\end{center}
%\end{table}

% COMPARISOM-SOTA: large testing set
\begin{table*}[t!]
	\centering
	\caption{Super-resolution of NIR images with the aid of RGB images. Performance comparison [in terms of average PSNR (dB)] over all test images for  $\times2$, $\times4$ and $\times6$ upscaling. }
	\label{tab:NIR_25}
	\begin{center}
		\addtolength{\tabcolsep}{0pt}
		\begin{tabular}{c || c | c | c | c | c |c | c | c || c  }
			\hline
			\hline
			{Scale}	&{CSCN~\cite{Huang}}	&{ACSC~\cite{raja}}	 &EDSR~\cite{EDSR}&SRFBN~\cite{SRFBN} &{DJF~\cite{DJF}}&{CoISTA~\cite{JMDL}}&LMCSC-Net& {LMCSC$_+$-Net}	&{LMCSC-ResNet}\\ 
			\hline
			$\times2$	&${36.84}$		&${36.92}$	&${36.59}$		&${36.94}$		&${38.78}$	&$38.93$	&$39.28$&$39.42$&${\bf{39.57}}$\\
			\hline
			$\times4$	&${30.64}$		&${30.35}$			&${31.85}$	&${32.11}$		&${32.91}$	&$33.20$	& 33.96&	34.23&${\bf{34.36}}$\\
			\hline
			$\times6$	&${27.94}$		&${27.91}$		&${30.49}$		&${30.63}$		&${29.19}$	&$30.88$	& 32.07&31.94 &${\bf{32.33}}$\\
			\hline	
			\hline
		\end{tabular}
	\end{center}
\end{table*}

%\begin{figure*}[t!]
%	\centering
%	\includegraphics[width=1\textwidth]{figures/dictionary.png}\\
%	\caption{The learned convolutional dictionary of LMCSC-ResNet. }
%	\label{fig:LMCSC-SR+}
%\end{figure*}

\begin{table*}[t!]
	\centering
	\caption{Super-resolution of NIR images with the aid of RGB images. Performance comparison  [in terms of PSNR (dB) and SSIM] for selected test images for  $\times2$, $\times4$ and $\times6$ upscaling. }
	\label{tab:NIR_8}
	\begin{center}

		\addtolength{\tabcolsep}{-4.1pt}
		\begin{tabular}{c || c  c | c c | c  c | c  c | c  c | c  c | c  c | c  c| c  c}
%			\cline{2-19}
 %           \hline			
			NIR/RGB&\multicolumn{2}{c|}{u-0004}&\multicolumn{2}{c|}{u-0006}&\multicolumn{2}{c|}{u-0017}&\multicolumn{2}{c|}{o-0018}&\multicolumn{2}{c|}{u-0020 }&\multicolumn{2}{c|}{u-0026}&\multicolumn{2}{c|}{o-0030}&\multicolumn{2}{c|}{u-0050}&\multicolumn{2}{c}{Average}\\ 
			\hline
			
			{$\times2$}&PSNR &SSIM & PSNR &SSIM &PSNR &SSIM &PSNR &SSIM &PSNR &SSIM &PSNR &SSIM &PSNR &SSIM &PSNR &SSIM&PSNR &SSIM
			\\
%			\hline
			\hline
			Bicubic &$ 30.55$&$ 0.9290$&$36.89$&$ 0.9462$&$ 34.83$&$ 0.9183$&$30.50$&$ 0.9116$&$32.68 $&$0.9324 $&$ 30.51$&$ 0.9142$&$30.90$&$0.8876$&$30.68$&$ 0.9229$&$32.19$&$ 0.9202$\\
			SDF~\cite{SDF} &$ 30.72$&$ 0.9290$&$ 36.71$&$ 0.9364$&$ 34.89$&$ 0.9139$&$ 30.74$&$ 0.9128$&$ 32.89$&$ 0.9317$&$ 30.58$&$ 0.9110$&$ 31.02$&$ 0.8816$&$ 30.61$&$ 0.9125$&$ 32.28$&$ 0.9161$\\
%			JBF~\cite{JBU} &$ 28.47$&$  0.9359$&$ 32.10$&$ 0.9311$&$  31.11$&$ 0.9172$&$ 27.59$&$ 0.9308$&$30.67 $&$0.9523 $&$ 26.82$&$ 0.8627$&$27.58$&$0.8476$&$27.32$&$ 0.9099$&$28.96$&$ 0.9109$\\
%			JFSM~\cite{JFSM} &$ 30.86$&$ 0.9721$&$ 32.86$&$ 0.9741$&$ 32.85 $&$ 0.9500$&$ 30.80$&$ 0.9774$&$32.61 $&$ 0.9797$&$ 28.97$&$ 0.9332$&$30.56$&$0.9064$&$27.58$&$ 0.9251$&$30.89$&$ 0.9522$\\
%			GF~\cite{MSR2} &$ 28.75$&$  0.9391$&$ 32.66$&$ 0.9400$&$ 31.32 $&$0.9205$&$ 27.70$&$ 0.9251$&$ 30.69$&$0.9494 $&$ 26.89$&$ 0.8571$&$27.59$&$0.8383$&$27.35$&$ 0.9116$&$29.12$&$ 0.9101$\\
			CSCN~\cite{Huang} &$ 32.77$&$ 0.9715$&$39.47$&$ 0.9715$&$ 36.76 $&$0.9574$&$ 33.98$&$0.9659$&$ 35.54$&$0.9658 $&$ 32.94$&$ 0.9339$&$33.34$&$0.9465$&$33.31$&$0.9693$&$34.76$&$ 0.9602$\\
			ACSC~\cite{raja} &$ 33.24$&$ 0.9723$&$39.78$&$ 0.9718$&$ 36.64 $&$0.9579$&$ 34.26$&$0.9670$&$ 35.65$&$0.9660 $&$ 33.11$&$ 0.9416$&$33.32$&$0.9465$&$33.39$&$0.9696$&$34.93$&$ 0.9608$\\
			EDSR~\cite{EDSR} &$ 35.10$&$ 0.9971$&$40.66$&$ 0.9958$&$ 37.57 $&$0.9942$&$ 35.97$&$0.9961$&$ 37.11$&$0.9968 $&$ 34.21$&$ 0.9941$&$34.82$&$0.9939$&$36.55$&$0.9956$&$36.49$&$ 0.9954$\\
			SRFBN~\cite{SRFBN}  &$ 35.36$&$ 0.9974$&$41.08$&$ 0.9970$&$ 38.19 $&$0.9950$&$ 36.47$&$0.9971$&$ 37.50$&$0.9969 $&$ 31.00$&$ 0.9782$&$35.57$&$0.9944$&$\textbf{37.06}$&$\textbf{0.9966}$&$36.53$&$ 0.9941$\\
%			MMSR~\cite{MMSR} &$ 37.55$&$  0.9934$&$ 45.21$&$ 0.9919$&$  39.68$&$ 0.9933$&$ 34.52$&$ 0.9912$&$38.87 $&$0.9883$&$ 37.50$&$ 0.9797$&$39.31$&$0.9943$&$38.95$&$ 0.9903$\\
			DJF~\cite{DJF} &$ 34.50$&$ 0.9964$&$41.52$&$ 0.9975$&$ 38.65 $&$0.9961
            $&$ 34.78$&$0.9960$&$ 37.35$&$0.9973 $&$ 33.15$&$   0.9939$&$35.67$&$0.9944$&$32.60$&$ 0.9928$&$36.03$&$0.9955$\\
%			CDLSR\cite{MSR} &$ \textbf{34.14}$&$  0.9811$&$ 36.79$&$ 0.9868$&$  35.27$&$ 0.9777$&$ 33.01$&$ 0.9874$&$36.66 $&$0.9893$&$ 30.35$&$ 0.9482$&$32.71$&$0.9443$&$29.37$&$ 0.9663$&$33.54$&$ 0.9726$\\
%			CoISTA~\cite{JMDL} &$31.56$&$  0.9835$&$ 37.21$&$ 0.9871$&$  34.87$&$ 0.9777$&$ 32.35$&$ 0.9867$&$34.75 $&$0.9887$&$ 29.94$&$ 0.9708$&$32.28$&$0.9709$&$29.42$&$ 0.9705$&$32.80$&$ 0.9795$\\
%			CU-Net~\cite{CUNet} &$ \textbf{38.35}$&$  -$&$ 42.76$&$ -$&$\textbf{38.74}$&$ -$&$\textbf{ 34.40}$&$-$&$\textbf{37.01 }$&$-$&$36.51$&$ -$&$36.89$&$-$&$\textbf{37.81}$&$ -$&$\textbf{37.81}$&$ -$\\
%			\hline
%			\hline
            CoISTA~\cite{JMDL} &$ 35.83$&$ 0.9968$&$ 42.40$&$ 0.9976$&$ 39.13$&$ 0.9953$&$ 37.54$&$ 0.9975$&$ 39.00$&$ 0.9974$&$ 34.11$&$ 0.9944$&$ 36.90$&$ 0.9946$&$ 33.53$&$ 0.9937$&$ 37.30$&$ 0.9959$\\
            DMSC~\cite{lesitaEUSIPCO}& $ 36.97$ & $0.9976 $ & $ 43.22$ &$ 0.9977$ &$ 40.41$ &$0.9970 $ &$ 37.90$ &$ 0.9964$ &$ 40.07$ &$ 0.9975$ &$ 34.96$ &$0.9948 $ &$ 37.74$ &$ 0.9953$ &$ 33.78$ &$0.9934 $ &$ 38.13$&$0.9962 $  \\
            
            \hline
            LMCSC-Net &$37.26 $&$ 0.9977$&$ 43.60$&$ \textbf{0.9982}$&$ 40.87$&$ 0.9967$&$ 39.21$&$ 0.9982$&$ 40.98$&$ \textbf{0.9980}$&$ 35.60$&$ \textbf{0.9963}$&$ \textbf{38.29}$&$ 0.9961$&$ 34.11$&$ 0.9948$&$ 38.74$&$ 0.9970$\\
            LMCSC$_+$-Net& $ 37.47$ & $0.9978 $ & $ 43.66$ &$0.9977 $ &$ \textbf{40.96}$ &$ \textbf{0.9988}$ &$ 39.58$ &$ 0.9975$ &$ 40.84$ &$ 0.9976$ &$ 35.92$ &$0.9950 $ &$ 38.14$ &$0.9959 $ &$ 34.27$ &$0.9940 $ &$ 38.84$&$0.9968 $  \\
			LMCSC-ResNet &$\textbf{37.99}$&$\textbf{0.9982}$&$\textbf{43.84}$&$0.9979$&$  40.76$&$0.9980$&$ \textbf{40.05}$&$\textbf{0.9989}$&$\textbf{41.05}$&$0.9978$&$ \textbf{36.15}$&$0.9954$&$38.18$&$\textbf{0.9965}$&$34.29$&$0.9943$&$\textbf{39.04}$&$\textbf{0.9971}$\\
			\hline
			\hline
						{$\times4$}&PSNR &SSIM & PSNR &SSIM &PSNR &SSIM &PSNR &SSIM &PSNR &SSIM &PSNR &SSIM &PSNR &SSIM &PSNR &SSIM&PSNR &SSIM
			\\
%			\hline
			\hline
			Bicubic &$ 25.93$&$ 0.9029$&$30.89$&$ 0.9458$&$ 30.45$&$ 0.9527$&$25.19 $&$ 0.9298$&$28.03 $&$0.9577 $&$ 26.27$&$ 0.8704$&$26.54$&$0.8401$&$26.65$&$ 0.9434$&$27.49$&$ 0.9179$\\
			SDF~\cite{SDF} &$ 26.82$&$  0.9066$&$ 30.60$&$ 0.8918$&$ 30.72$&$ 0.9281$&$ 26.09$&$ 0.9169$&$29.09 $&$0.9505 $&$ 26.61$&$ 0.8558$&$27.21$&$0.8415$&$27.07$&$0.9207$&$28.03$&$0.9018$\\
			JBF~\cite{JBU} &$ 28.47$&$  0.9359$&$ 32.10$&$ 0.9311$&$  31.11$&$ 0.9172$&$ 27.59$&$ 0.9308$&$30.67 $&$0.9523 $&$ 26.82$&$ 0.8627$&$27.58$&$0.8476$&$27.32$&$ 0.9099$&$28.96$&$ 0.9109$\\
			JFSM~\cite{JFSM} &$ 30.86$&$ 0.9721$&$ 32.86$&$ 0.9741$&$ 32.85 $&$ 0.9500$&$ 30.80$&$ 0.9774$&$32.61 $&$ 0.9797$&$ 28.97$&$ 0.9332$&$30.56$&$0.9064$&$27.58$&$ 0.9251$&$30.89$&$ 0.9522$\\
			GF~\cite{MSR2} &$ 28.75$&$  0.9391$&$ 32.66$&$ 0.9400$&$ 31.32 $&$0.9205$&$ 27.70$&$ 0.9251$&$ 30.69$&$0.9494 $&$ 26.89$&$ 0.8571$&$27.59$&$0.8383$&$27.35$&$ 0.9116$&$29.12$&$ 0.9101$\\
			CSCN~\cite{Huang} &$ 27.64$&$  0.9378$&$ 32.60$&$ 0.9361$&$ 31.68 $&$0.9211$&$ 27.28$&$ 0.9250$&$ 30.04$&$0.9487 $&$ 27.91$&$ 0.8724$&$27.72$&$0.8378$&$28.20$&$ 0.9101$&$29.14$&$ 0.9111$\\
			ACSC~\cite{raja} &$ 27.28$&$  0.9371$&$ 32.61$&$ 0.9360$&$ 31.66 $&$0.9208$&$ 27.42$&$ 0.9252$&$ 29.87$&$0.9483 $&$ 27.92$&$ 0.8722$&$27.66$&$0.8381$&$27.80$&$ 0.9086$&$29.03$&$ 0.9107$\\
			EDSR~\cite{EDSR} &$ 28.59$&$ 0.9759$&$ 33.42$&$ 0.9684$&$  32.50$&$ 0.9693$&$ 28.54$&$ 0.9717$&$31.09$&$ 0.9789$&$ 28.74$&$ 0.9678$&$28.81$&$0.9528$&$29.58$&$ 0.9742$&$30.15$&$ 0.9699$\\
			SRFBN~\cite{SRFBN}  &$ 29.01$&$ 0.9787$&$ 33.73$&$ 0.9702$&$  32.91$&$ 0.9725$&$ 28.88$&$ 0.9740$&$31.44$&$0.9807 $&$ 29.10$&$ 0.9702$&$29.45$&$0.9583$&$29.89$&$ 0.9762$&$30.55$&$ 0.9726$\\
%			MMSR~\cite{MMSR} &$ 37.55$&$  0.9934$&$ 45.21$&$ 0.9919$&$  39.68$&$ 0.9933$&$ 34.52$&$ 0.9912$&$38.87 $&$0.9883$&$ 37.50$&$ 0.9797$&$39.31$&$0.9943$&$38.95$&$ 0.9903$\\
			DJF~\cite{DJF} &$ 31.02$&$  0.9784$&$ 36.04$&$ 0.9894$&$  34.18$&$ 0.9815$&$ 30.72$&$ 0.9888$&$33.60 $&$0.9915$&$ 29.21$&$ 0.9397$&$31.27$&$0.9345$&$28.58$&$ 0.9616$&$31.83$&$ 0.9707$\\
%			CDLSR\cite{MSR} &$ \textbf{34.14}$&$  0.9891$&$ 36.79$&$ 0.9868$&$  35.27$&$ 0.9777$&$ 33.01$&$ 0.9874$&$36.66 $&$0.9893$&$ 30.35$&$ 0.9482$&$32.71$&$0.9443$&$29.37$&$ 0.9663$&$33.54$&$ 0.9726$\\
			CoISTA~\cite{JMDL} &$31.56$&$  0.9835$&$ 37.21$&$ 0.9871$&$  34.87$&$ 0.9777$&$ 32.35$&$ 0.9867$&$34.75 $&$0.9887$&$ 29.94$&$ 0.9708$&$32.28$&$0.9709$&$29.42$&$ 0.9705$&$32.80$&$ 0.9795$\\
%			CU-Net~\cite{CUNet} &$ \textbf{38.35}$&$  -$&$ 42.76$&$ -$&$\textbf{38.74}$&$ -$&$\textbf{ 34.40}$&$-$&$\textbf{37.01 }$&$-$&$36.51$&$ -$&$36.89$&$-$&$\textbf{37.81}$&$ -$&$\textbf{37.81}$&$ -$\\
%			\hline
%			\hline
            DMSC~\cite{lesitaEUSIPCO}& $ 33.19$ & $0.9846 $ & $ 37.69$ &$ 0.9892$ &$ 36.00$ &$ 0.9812$ &$ 33.84$ &$ 0.9888$ &$ 36.33$ &$ 0.9893$ &$ 30.65$ &$ 0.9725$ &$ 33.19$ &$0.9727 $ &$ 29.85$ &$0.9716 $ &$ 33.84$&$0.9812 $  \\
            \hline
            LMCSC-Net & $ 33.75$ & $ 0.9869$ & $ 38.74$ &$0.9912 $ &$ 36.16$ &$0.9828 $ &$ 34.17$ &$ 0.9902$ &$ 36.95$ &$ 0.9900$ &$ 31.03$ &$ 0.9784$ &$ 33.56$ &$ 0.9780$ &$ 30.04$ &$ 0.9772$ &$ 34.28$&$0.9843 $  \\
            LMCSC$_+$-Net& $ \textbf{33.89}$ & $ 0.9889$ & $ 38.69$ &$ 0.9909$ &$ 36.13$ &$ 0.9828$ &$ 34.54$ &$ 0.9914$ &$ 37.08$ &$0.9909 $ &$ \textbf{31.21}$ &$\textbf{0.9785} $ &$ \textbf{33.66}$ &$ 0.9784$ &$ 30.07$ &$0.9771 $ &$ 34.39$&$ 0.9889$  \\
			LMCSC-ResNet &$33.80$&$\textbf{0.9895}$&$\textbf{38.77}$&$\textbf{0.9915}$&$  \textbf{36.54}$&$\textbf{0.9836}$&$ \textbf{34.78}$&$\textbf{0.9920}$&$\textbf{37.34}$&$\textbf{0.9923}$&$ 30.95$&$0.9782$&$33.59$&$\textbf{0.9784}$&$\textbf{30.10}$&$\textbf{0.9773}$&$\textbf{34.49}$&$\textbf{0.9853}$\\
			\hline
			\hline
%			&\multicolumn{2}{c||}{u-0004}&\multicolumn{2}{c||}{u-0006}&\multicolumn{2}{c||}{u-0017}&\multicolumn{2}{c||}{o-0018}&\multicolumn{2}{c||}{u-0020 }&\multicolumn{2}{c||}{u-0026}&\multicolumn{2}{c||}{o-0030}&\multicolumn{2}{c||}{u-0050}&\multicolumn{2}{c}{Average}\\ \cline{2-19}
			
			{$\times6$}&PSNR &SSIM & PSNR &SSIM &PSNR &SSIM &PSNR &SSIM &PSNR &SSIM &PSNR &SSIM &PSNR &SSIM &PSNR &SSIM&PSNR &SSIM
			\\
			\hline
%			\hline
			Bicubic &$ 23.87$&$ 0.8094$&$28.48$&$ 0.8671$&$ 28.64$&$ 0.8998$&$23.07 $&$ 0.8393$&$26.03 $&$0.9053 $&$ 24.71$&$ 0.7850$&$25.19$&$0.7515$&$25.17$&$ 0.8921$&$25.65$&$ 0.8437$\\
			SDF~\cite{SDF} &$ 24.62$&$  0.8413$&$ 28.76$&$ 0.8377$&$ 29.13$&$ 0.8910$&$ 23.79$&$ 0.8439$&$26.93 $&$0.9089 $&$ 25.17$&$ 0.7989$&$25.80$&$0.7748$&$25.90$&$0.8837$&$26.26$&$0.8475$\\
			JBF~\cite{JBU} &$ 25.96$&$  0.8858$&$ 30.00$&$ 0.8861$&$ 29.63$&$ 0.8864$&$ 25.09$&$ 0.8718$&$28.19 $&$0.9200 $&$ 25.64$&$ 0.8235$&$26.32$&$0.7994$&$26.26$&$ 0.8837$&$27.13$&$ 0.8696$\\
			JFSM~\cite{JFSM} &$ 27.97$&$ 0.9527$&$ 32.28$&$ 0.9716$&$ 32.01 $&$ 0.9434$&$ 27.47$&$ 0.9470$&$30.33 $&$ 0.9673$&$ 27.54$&$ 0.9128$&$29.38$&$0.8902$&$26.67$&$ 0.9068$&$29.21$&$ 0.9365$\\
			GF~\cite{MSR2} &$ 25.94$&$  0.8817$&$ 30.17$&$ 0.8876$&$  29.61$&$0.8860$&$ 24.98$&$ 0.8591$&$ 28.01$&$0.9118 $&$ 25.63$&$ 0.8131$&$26.22$&$0.7855$&$26.26$&$ 0.8846$&$27.10$&$ 0.8637$\\
			CSCN~\cite{Huang} &$ 25.01$&$ 0.8796$&$ 29.94$&$ 0.8860$&$ 29.53 $&$ 0.8788$&$ 24.57$&$ 0.8521$&$27.46 $&$ 0.9105$&$ 25.79$&$ 0.8052$&$25.86$&$0.7698$&$26.71$&$ 0.9026$&$26.86$&$ 0.8605$\\
			ACSC~\cite{raja} &$ 24.82$&$ 0.8790$&$ 29.97$&$ 0.8856$&$ 29.48$&$ 0.8789$&$ 24.70$&$ 0.8508$&$27.49 $&$ 0.9111$&$ 25.97$&$ 0.8049$&$25.91$&$0.7704$&$26.43$&$ 0.9031$&$26.85$&$ 0.8604$\\
			EDSR~\cite{EDSR} &$ 26.52$&$ 0.9536$&$ 30.60$&$0.9340 $&$  30.55$&$ 0.9470$&$ 25.67$&$0.9379 $&$28.56$&$ 0.9590$&$ 26.96$&$ 0.9408$&$26.76$&$0.9182$&$27.58$&$ 0.9508$&$27.90$&$ 0.9426$\\
			SRFBN~\cite{SRFBN}  &$ 26.62$&$  0.9559$&$ 30.83$&$ 0.9342$&$  30.80$&$ 0.9500$&$ 25.93$&$ 0.9385$&$28.64$&$0.9591 $&$ 27.19$&$ 0.9422$&$27.24$&$0.9225$&$27.58$&$ 0.9502$&$28.10$&$ 0.9441$\\
%			MMSR~\cite{MMSR} &$ 37.55$&$  0.9934$&$ 45.21$&$ 0.9919$&$  39.68$&$ 0.9933$&$ 34.52$&$ 0.9912$&$38.87 $&$0.9883$&$ 37.50$&$ 0.9797$&$39.31$&$0.9943$&$38.95$&$ 0.9903$\\
			DJF~\cite{DJF} &$ 29.68$&$  0.9670$&$ 34.92$&$ 0.9830$&$  32.80$&$ 0.9599$&$ 29.92$&$ 0.9844$&$32.61 $&$0.9873$&$ 28.38$&$ 0.9183$&$30.00$&$0.9063$&$27.64$&$ 0.9414$&$30.75$&$ 0.9559$\\
%			CDLSR\cite{MSR} &$ 30.77$&$  0.9558$&$ 34.15$&$ 0.9664$&$  32.98$&$ 0.9515$&$ 31.03$&$ 0.9727$&$33.85 $&$0.9763$&$ 28.88$&$ 0.9172$&$30.52$&$0.9099$&$28.37$&$ 0.9402$&$31.32$&$ 0.9487$\\
			CoISTA~\cite{JMDL} &$29.19$&$  0.9686$&$ 34.82$&$ 0.9769$&$  32.79$&$ 0.9549$&$29.94$&$ 0.9755$&$32.38 $&$0.9791$&$ 28.53$&$ 0.9487$&$30.22$&$0.9479$&$28.30$&$ 0.9539$&$30.78$&$ 0.9632$\\
%			CU-Net~\cite{CUNet} &$ \textbf{38.35}$&$  -$&$ 42.76$&$ -$&$\textbf{38.74}$&$ -$&$\textbf{34.40}$&$-$&$\textbf{37.01}$&$-$&$36.51$&$ -$&$36.89$&$-$&$\textbf{37.81}$&$ -$&$\textbf{37.81}$&$ -$\\
%			\hline
%			\hline
            DMSC~\cite{lesitaEUSIPCO}& $ 30.20$ & $ 0.9734$ & $ 35.74$ &$ 0.9815$ &$ 33.55$ &$ 0.9778$ &$ 31.34$ &$ 0.9796$ &$ 33.56$ &$ 0.9822$ &$ 29.01$ &$ 0.9603$ &$ 30.61$ &$0.9584 $ &$ 28.45$ &$ 0.9591$ &$ 31.56$&$ 0.9715$  \\
            \hline
            LMCSC-Net& $ 31.74$ & $ 0.9811$ & $ 36.46$ &$ 0.9851$ &$ 34.23$ &$0.9748 $ &$ 32.88$ &$ 0.9865$ &$ 35.23$ &$0.9872 $ &$ 29.37$ &$ 0.9622$ &$ 31.08$ &$0.9599 $ &$ 28.74$ &$ 0.9609$ &$ 32.46$&$0.9751 $  \\
            LMCSC$_+$-Net& $ 32.12$ & $0.9840 $ & $ 36.45$ &$0.9850 $ &$ 34.58$ &$0.9759 $ &$ 33.17$ &$ 0.9873$ &$ 35.68$ &$ 0.9888$ &$ 29.51$ &$0.9628 $ &$ 31.53$ &$ 0.9631$ &$ 28.89$ &$ 0.9622$ &$ 32.74$&$0.9762 $  \\
			LMCSC-ResNet &$\textbf{32.19}$&$\textbf{0.9846}$&$\textbf{36.52}$&$\textbf{0.9860}$&$  \textbf{34.65}$&$\textbf{0.9763}$&$ \textbf{33.24}$&$\textbf{0.9877}$&$\textbf{35.75}$&$\textbf{0.9889}$&$ \textbf{29.58}$&$\textbf{0.9630}$&$\textbf{31.59}$&$\textbf{0.9638}$&$\textbf{28.94}$&$\textbf{0.9624}$&$\textbf{32.81}$&$\textbf{0.9766}$\\
			\hline
			\hline
		\end{tabular}
	\end{center}
\end{table*}

We apply the proposed models to different upsampling tasks, 
that is, super-resolution of near-infrared (NIR) images, depth upsampling and super-resolution of multi-spectral data,
using RGB images as side information.
We compare our models against state-of-the-art single-modal and multimodal methods
showing the superior performance of the proposed approach.
Before demonstrating our experimental results,
we present the employed datasets
and report implementation details.

\subsection{Datasets}
\label{sec:dataset}
\subsubsection{EPFL RGB-NIR dataset} 
NIR images are acquired at a low resolution 
due to the high cost per pixel of a NIR sensor
compared to an RGB sensor.
We employ the EPFL RGB-NIR dataset\footnote{https://ivrl.epfl.ch/supplementary\_material/cvpr11/} 
and apply our models to super-resolve an LR NIR image with the aid of an HR RGB image.
The dataset contains $477$ spatially aligned NIR/RGB  image pairs.
Our training set contains approximately $30,000$ cropped image pairs
extracted from $50$ images.
Each training image is of size $44 \times 44$ pixels; 
the size is chosen with respect to
memory requirements and computational complexity. 
We also create a testing set containing $25$ image pairs;
testing is performed on an entire image.

The NIR images consist of one channel.
An LR version of a NIR image is generated by blurring and downscaling the ground truth HR version. 
We convert the RGB images to YCbCr 
and only utilize the luminance channel as the side information. 
Following~\cite{dong2016image},
we apply bicubic interpolation as a preprocessing step to upscale the LR input 
such that the input and output images are of the same size.

\subsubsection{NYU v2 RGB-D dataset}
Depth cameras like Microsoft Kinect and time-of-flight (ToF) cameras only provide low-resolution depth images. 
Therefore, depth upsampling is a necessary task for many vision applications. 
We apply our models for depth upsampling with the aid of RGB images,
using the NYU v2 RGB-D dataset~\cite{NYU}.
The dataset contains $1449$ RGB images with their depth maps.
Similar to~\cite{DJF}, we use the first $1000$ images for training, 
and the remaining $449$ for testing. 

\subsubsection{Columbia multi-spectral database}
The third dataset that we use to evaluate the proposed models 
is the Columbia multi-spectral database,\footnote{http://www.cs.columbia.edu/CAVE/databases/multispectral}
which contains spectral reflectance data and RGB images. 
For testing, we reserve $7$ images from the $640$~nm band 
and randomly select $7$ images  from different bands;
the rest are used for training.  

%%%%%%%%%%%%%%%%%%%%%%%%%%%%%%%%%%
\subsection{Implementation Details}
\label{sec:parameters}
%%%%%%%%%%%%%%%%%%%%%%%%%%%%%%%%%%
All networks are designed with three unfolding steps 
for the target (LMCSC) and the side information (ACSC) encoders.
The number of unfolding steps is chosen after taking into account  
the trade-off between the computational complexity and the reconstruction accuracy;
for instance, by increasing the unfolding steps to five, 
the improvement in the average PSNR is less than $0.1$~dB 
while the execution time is almost $87\%$ higher. 
In the LMCSC-ResNet, the ACSC branch employed for the nonlinear mapping of the target signal
is designed with one unfolding step. 

We empirically set the size of the network parameters 
$\boldsymbol{\mathcal{P}}$, $\boldsymbol{\mathcal{Q}}$, $\boldsymbol{\mathcal{G}}$ and $\boldsymbol{\mathcal{T}}$ to $7\times7\times1\times85$;
the size of $\boldsymbol{\mathcal{R}}$ and $\boldsymbol{\mathcal{V}}$ are set to $7 \times 7 \times 85 \times 1$. 
The size of the convolutional dictionaries for reconstruction is $7\times 7 \times 85 \times 1$. 
Note that a convolutional layer of size $k\times k\times c\times g$ 
consists of $g$ convolutional filters with kernel size $k$ and $c$ channels. 
We use untied weights at every unfolding step, i.e., 
the $t$-th layer of LMCSC and ACSC subnetworks is realized 
by the independent variables $\boldsymbol{\mathcal{R}}_t$, $\boldsymbol{\mathcal{Q}}_t$ and $\boldsymbol{\mathcal{V}}_t$, $\boldsymbol{\mathcal{T}}_t$, respectively.
The weights of all layers are initialized randomly 
using the Gaussian distribution with standard deviation equal to $0.01$. 
The parameters $\mu$ and $\gamma$ of the proximal operators 
are both initialized to $0.2$. 
We train the network using the Adam optimizer.

We notice that the complexity of our networks is dominated by the LeSITA activation layers in the LMCSC block, and an implementation based on~\eqref{eq:prox1} and~\eqref{eq:prox2} is not efficient. In order to address this issue, we rewrite the proximal operator in~\eqref{eq:prox1},~\eqref{eq:prox2} as follows:
\begin{equation}
\label{eq:LeSITA_new}
\begin{split}
\xi_{\mu}(v_i;s_i) =& \sign(s_i)[R(\sign(s_i)v_i-2\mu-|s_i|)\\
&-R(\sign(s_i)v_i-|s_i|)+R(\sign(s_i)v_i)\\
&-R(-\sign(s_i)v_i-2\mu)],
\end{split}
\end{equation}
where $R(v)=\max(0,v)$ is the Rectified Linear Unit (ReLU) function. 
This form of the proximal operator results in a $30$\% faster implementation and we use this version in all of the experiments.

%%%%%%%%%%%%%%%%%%%%%%%%%%%%%%%%%%
\subsection{Performance Comparison}
\label{sec:compare}
%%%%%%%%%%%%%%%%%%%%%%%%%%%%%%%%%%
%
\begin{table*}
	\small
	\centering
	\caption{Depth upsampling with the aid of RGB images. Performance comparison  [in terms of average RMSE] over $449$ test images from the NYU v2 RGB-D dataset
	for $\times 4$, $\times 8$, and $\times 16$ upsampling. }
	\label{tab:NYU_449}
	\begin{center}
		\addtolength{\tabcolsep}{0pt}
		\begin{tabular}{  c| c|  c| c|  c| c|| c | c | c }
			\hline
			\hline
			
			RGB-D&GF~\cite{MSR2}&JBF~\cite{JBU}& SDF~\cite{SDF}&DJF\cite{DJF}&Gu \textit{et al.}~\cite{dynamic-guidance}& LMCSC-Net& LMCSC$_+$-Net &LMCSC-ResNet\\ 
			
			\hline
			%			\hline
			$\times4$  & 4.04 & 2.31 & 3.04 & 1.97 & 1.56 & 1.49& \textbf{1.38}& 1.45 \\
			\hline
			
			$\times8$  & 7.34 & 4.12 & 5.67 & 3.39 & 2.99 & 2.67& \textbf{2.58}& 2.61 \\
			\hline
			
			$\times16$ & 12.23 & 6.98 & 9.97 & 5.63 & 5.24 &5.01 &4.93 & \textbf{4.88}\\
			\hline
			\hline
			
		\end{tabular}
	\end{center}
\end{table*}

\begin{table*}[t!]
	\centering
	\caption{Depth upsampling with the aid of RGB images. 
	Performance comparison  [in terms of RMSE] over selected test images from the NYU v2 dataset for $\times 4$, $\times 8$ and $\times16$ upsampling. }
	\label{tab:NYU_10}
	\begin{center}
		\addtolength{\tabcolsep}{0.7pt}
		\begin{tabular}{c | c  ||   c   c    c    c    c   c    c   c  c  c  c }
%			\hline
			
			\multicolumn{2}{c||}{Depth/RGB}&\multicolumn{1}{c}{NYU-1}&\multicolumn{1}{c}{NYU-2}&\multicolumn{1}{c}{NYU-3}&\multicolumn{1}{c}{NYU-4}&\multicolumn{1}{c}{NYU-5}&\multicolumn{1}{c}{NYU-6}&\multicolumn{1}{c}{NYU-7}&\multicolumn{1}{c}{NYU-8}&\multicolumn{1}{c}{NYU-9}&\multicolumn{1}{c}{NYU-10}&\multicolumn{1}{c}{Average}\\ 
			\hline

			\hline
%			\hline
			\multirow{6}{*}{$\times4$}&Bicubic &$ 3.70$&$ 4.35$&$2.77$&$ 3.73$&$ 4.13$&$ 4.43$&$4.31 $&$ 3.52$&$3.33 $&$5.20 $&$ 3.95$\\
			&DJF~\cite{DJF} &$ 1.63$&$ 2.47$&$1.31$&$ 2.08$&$ 2.07$&$ 1.77$&$1.80 $&$ 1.73$&$1.55 $&$2.25 $&$ 1.87$\\
			&Gu \textit{et al.}~\cite{dynamic-guidance} &$ 1.59 $&$1.53 $&$1.08 $&$1.62  $&$1.78 $&$ 1.62 $&$ 2.17$&$ 1.57 $&$ 1.43 $&$ 2.28$&$ 1.66$\\
			&LMCSC-Net &$ 1.38$&$ 1.08$&$ 0.96$&$ 1.05$&$ 1.57$&$ 1.28$&$ 1.31$&$ 1.29$&$ 1.13$&$ 1.66$&$ 1.28$\\
			&LMCSC$_+$-Net &$\textbf{1.35}$&$ \textbf{1.02}$&$ \textbf{0.88}$&$ \textbf{1.04}$&$ \textbf{1.45}$&$ \textbf{1.16}$&$ \textbf{1.28}$&$ \textbf{1.22}$&$ \textbf{1.08}$&$ \textbf{1.39}$&$ \textbf{1.19}$\\
        	&LMCSC-ResNet &$ 1.37$&$ 1.13$&$0.92$&$ 1.10$&$1.50$&$ 1.26$&$1.42 $&$ 1.24$&$1.09 $&$1.51 $&$ 1.26$\\

			\hline
%			\hline

			\hline
%			\hline
			\multirow{6}{*}{$\times8$}&Bicubic &$ 6.42$&$ 6.16$&$4.83$&$ 6.93$&$ 6.94$&$ 7.07$&$7.19$&$ 6.11$&$5.91 $&$8.63 $&$ 6.62$\\
			&DJF~\cite{DJF} &$ 3.04$&$ 3.10$&$2.22$&$ 3.96$&$ 3.51$&$ 2.84$&$3.00 $&$ 2.92$&$2.75 $&$4.41 $&$ 3.17$\\
			&Gu \textit{et al.}~\cite{dynamic-guidance} &$ 2.93 $&$3.55 $&$2.12 $&$ 3.47 $&$ 2.95$&$ 2.87 $&$3.80 $&$2.95  $&$ 2.50 $&$4.19 $&$ 3.13$\\
			&LMCSC-Net &$ 2.16$&$ 2.01$&$ 1.47$&$ 2.33$&$ 2.95$&$ 2.18$&$ 2.37$&$ 2.23$&$ 1.81$&$ 2.80$&$ 2.23$\\
			&LMCSC$_+$-Net &$ 2.14$&$ 1.96$&$ 1.47$&$ 2.30$&$ \textbf{2.86}$&$ \textbf{2.14}$&$ \textbf{2.29}$&$ 2.23$&$ 1.78$&$ 2.84$&$ 2.20$\\
			&LMCSC-ResNet &$ \textbf{2.09}$&$ \textbf{1.93}$&$\textbf{1.42}$&$ \textbf{2.21}$&$ 2.90$&$ 2.15$&$2.40 $&$ \textbf{2.22}$&$\textbf{1.75} $&$\textbf{2.66} $&$ \textbf{2.18}$\\

			\hline
%			\hline

			\hline
%			\hline
			\multirow{6}{*}{$\times16$}&Bicubic &$ 11.10$&$ 11.99$&$7.88$&$ 10.49$&$ 11.04$&$ 11.59$&$11.11 $&$ 10.55$&$9.86 $&$14.67 $&$ 11.03$\\
			&DJF~\cite{DJF} &$ 5.20$&$ 4.23$&$3.53$&$ 6.21$&$ 6.10$&$ 3.78$&$5.30 $&$ 5.00$&$ 4.42$&$6.72 $&$ 5.05$\\
			&Gu \textit{et al.}~\cite{dynamic-guidance} &$  4.95$&$ 5.15$&$ 3.34$&$  5.61$&$ 5.12$&$4.20  $&$5.45 $&$ 4.66 $&$ 4.18 $&$6.80 $&$ 4.95$\\
			&LMCSC-Net &$ 3.86$&$ 4.37$&$ 3.55$&$ \textbf{4.59}$&$ 5.53$&$ 4.18$&$ 4.53$&$ 3.86$&$ 3.61$&$ 6.03$&$ 4.41$\\
			&LMCSC$_+$-Net &$ \textbf{3.80}$&$ 4.12$&$ 3.32$&$ 4.61$&$ 5.48$&$ 4.01$&$ 4.23$&$ \textbf{3.84}$&$ 3.55$&$ \textbf{5.82}$&$ 4.29$\\
			&LMCSC-ResNet &$ 3.94$&$ \textbf{3.89}$&$\textbf{3.16}$&$ 4.85$&$ \textbf{4.73}$&$ \textbf{3.54}$&$\textbf{4.18} $&$ 3.87$&$\textbf{3.37} $&$5.87$&$ \textbf{4.15}$\\

			\hline
			\hline
		\end{tabular}
	\end{center}
\end{table*}

%\begin{table*}
%	\small
%	\centering
%	\caption{SR results on Columbia multispectral database [in terms of SSIM and PSNR]. Test images are chosen from 640 nm band. }
%	\label{tab:table5}
%	\begin{center}
%		\addtolength{\tabcolsep}{-1pt}
%		\begin{tabular}{c | c c | c c |  c c | c c | c c | c c || c c  }
%			\hline
%			\hline
%			\multirow{2}{*}{scale}&\multicolumn{2}{c|}{JBU~\cite{JBU}}&\multicolumn{2}{c|}{SDF~\cite{SDF} }&\multicolumn{2}{c|}{JFSM~\cite{JFSM} }&\multicolumn{2}{c|}{DJF\cite{DJF} }&\multicolumn{2}{c|}{CDLSR\cite{MSR}}&\multicolumn{2}{c||}{CoISTA\cite{JMDL}}&\multicolumn{2}{c}{\textbf{LMCSC-ResNet}}\\ 
%			&SSIM &PSNR&SSIM &PSNR&SSIM &PSNR&SSIM &PSNR&SSIM &PSNR&SSIM &PSNR&SSIM &PSNR\\
%			%			
%			\hline
%			%			\hline
%			$\times4$ & 0.9265 & 31.17 & 0.9123 & 30.63 & 0.9379 &33.64&0.9758 & 34.94&0.9791& 35.60& - & 36.24& \textbf{0.9905}& \textbf{36.78} \\
%			\hline
%			
%			$\times6$ &0.8704 & 28.76 & 0.8591 & 28.50 & 0.9330 &32.00&0.9638 &33.62&0.9642 & 33.93& - & -& \textbf{0.9824}& \textbf{34.87} \\
%			\hline
%			\hline
%			
%		\end{tabular}
%	\end{center}
%\end{table*}

\begin{table*}[t!]
	\centering
	\caption{Super-resolution of multi-spectral images with the aid of RGB images. Performance comparison [in terms of PSNR (dB) and SSIM] for selected multi-spectral test images from 640nm band for $\times4$ upsampling. }
	\label{tab:MS_640nm}
	\begin{center}
		\addtolength{\tabcolsep}{-3pt}
		\begin{tabular}{c || c  c | c  c | c  c | c  c | c  c | c  c | c  c | c c}
			\hline
			
			\multirow{2}{*}{$\times4$}&\multicolumn{2}{c|}{Chart toy}&\multicolumn{2}{c|}{Egyptian}&\multicolumn{2}{c|}{Feathers}&\multicolumn{2}{c|}{Glass tiles}&\multicolumn{2}{c|}{Jelly beans}&\multicolumn{2}{c|}{Oil Paintings}&\multicolumn{2}{c|}{Paints}&\multicolumn{2}{c}{Average}\\ \cline{2-17}
			
			&PSNR &SSIM & PSNR &SSIM &PSNR &SSIM &PSNR &SSIM &PSNR &SSIM &PSNR &SSIM &PSNR &SSIM &PSNR &SSIM
			\\
			\hline
%			\hline
			Bicubic &$ 29.14$&$ 0.9451$&$36.22$&$ 0.9761$&$ 30.46$&$ 0.9530$&$26.38 $&$ 0.92156$&$27.45 $&$0.9269 $&$ 32.23$&$ 0.9025$&$30.47$&$0.9569$&$30.39$&$ 0.9403$\\
			SDF~\cite{SDF} &$ 30.74$&$  0.9523$&$ 37.16$&$ 0.9677$&$ 30.92$&$ 0.9434$&$ 27.01$&$ 0.9188$&$27.87 $&$0.9279 $&$ 32.80$&$ 0.9001$&$31.35$&$0.9569$&$31.12$&$0.9381$\\
			JBF~\cite{JBU} &$ 30.69$&$  0.9528$&$ 37.82$&$ 0.9788$&$  31.80$&$ 0.9599$&$ 27.15$&$ 0.9339$&$28.97 $&$0.9474 $&$ 33.23$&$ 0.9034$&$32.08$&$0.9714$&$31.68$&$ 0.9497$\\
			JFSM~\cite{JFSM} &$ 33.30$&$ 0.9215$&$ 39.68$&$ 0.9428$&$ 33.54 $&$ 0.9096$&$ 29.34$&$ 0.9407$&$30.82 $&$ 0.9356$&$ 34.16$&$ 0.9439$&$32.96$&$0.9321$&$33.40$&$ 0.9323$\\
			GF~\cite{MSR2} &$ 30.70$&$  0.9514$&$ 37.96$&$ 0.9788$&$  32.12$&$0.9618$&$ 27.45$&$ 0.9326$&$ 29.54$&$0.9488 $&$ 33.30$&$ 0.9033$&$32.23$&$0.9698$&$31.90$&$ 0.9494$\\
			EDSR~\cite{EDSR} &$ 33.57$&$ 0.9926$&$ 39.42$&$ 0.9941$&$  35.11$&$ 0.9927$&$ 29.41$&$ 0.9822$&$32.67$&$0.9917 $&$ 33.40$&$ 0.9676$&$37.33$&$0.9955$&$34.42$&$ 0.9880$\\
			SRFBN~\cite{SRFBN}  &$ 33.61$&$  0.9925$&$ 39.44$&$ 0.9942$&$  35.10$&$ 0.9933$&$ 29.35$&$ 0.9823$&$32.74$&$0.9919 $&$ 33.42$&$ 0.9677$&$37.23$&$0.9957$&$34.41$&$ 0.9882$\\
%			MMSR~\cite{MMSR} &$ 37.55$&$  0.9934$&$ 45.21$&$ 0.9919$&$  39.68$&$ 0.9933$&$ 34.52$&$ 0.9912$&$38.87 $&$0.9883$&$ 37.50$&$ 0.9797$&$39.31$&$0.9943$&$38.95$&$ 0.9903$\\
			DJF~\cite{DJF} &$ 33.91$&$  0.9842$&$ 41.31$&$ 0.9861$&$  36.01$&$ 0.9848$&$ 31.83$&$ 0.9814$&$32.77 $&$0.9820$&$ 34.39$&$ 0.9493$&$37.74$&$0.9897$&$35.43$&$ 0.9796$\\
%			CDLSR\cite{MSR} &$ 35.79$&$  0.9875$&$ 42.62$&$ 0.9938$&$  37.25$&$ 0.9887$&$ 32.26$&$ 0.9828$&$35.72 $&$0.9886$&$ 36.27$&$ 0.9556$&$37.24$&$0.9907$&$36.74$&$ 0.9839$\\
			CoISTA~\cite{JMDL} &$36.63$&$  0.9937$&$ 41.19$&$ 0.9948$&$  36.42$&$ 0.9913$&$ 30.63$&$ 0.9848$&$34.37 $&$0.9919$&$ 35.43$&$ 0.9791$&$34.84$&$0.9928$&$35.64$&$ 0.9898$\\
%			CU-Net~\cite{CUNet} &$ \textbf{38.35}$&$  \textbf{0.9962}$&$ 42.76$&$ \textbf{0.9960}$&$\textbf{38.74}$&$ \textbf{0.9955}$&$\textbf{34.40}$&$\textbf{0.9945}$&$\textbf{37.01}$&$\textbf{0.9961}$&$36.51$&$ 0.9845$&$36.89$&$0.9960$&$\textbf{37.81}$&$ \textbf{0.9941}$\\
%			\hline
			\hline
			LMCSC-Net &$ \textbf{38.47}$&$ \textbf{0.9962}$&$ 42.47$&$ 0.9956$&$ \textbf{38.32}$&$ \textbf{0.9953}$&$ 33.52$&$ 0.9936$&$ 36.75$&$ 0.9955$&$ 36.78$&$ 0.9844$&$ 36.44$&$ 0.9955$&$ 37.53$&$ 0.9937$\\
			LMCSC$_+$-Net &$ 38.01$&$ 0.9959$&$ 43.10$&$ 0.9958$&$ 37.94$&$0.9950$&$ \textbf{33.73}$&$ \textbf{0.9936}$&$ \textbf{36.89}$&$ 0.9956$&$ 36.32$&$ 0.9838$&$ 36.28$&$ 0.9956$&$ 37.47$&$ 0.9936$\\ 
			LMCSC-ResNet &$38.30$&$0.9958$&$\textbf{43.45}$&$\textbf{0.9958}$&$  38.14$&$0.9949$&$ 33.72$&$0.9932$&$36.83$&$\textbf{0.9956}$&$ \textbf{36.94}$&$\textbf{0.9846}$&$\textbf{36.95}$&$\textbf{0.9961}$&$\textbf{37.76}$&$\textbf{0.9937}$\\
			\hline
			\hline
		\end{tabular}
	\end{center}
\end{table*}

\begin{table*}[t!]
	\centering
	\caption{Super-resolution of multi-spectral images with the aid of RGB images. 
	Performance comparison [in terms of PSNR (dB) and SSIM] over selected multi-spectral test images (from different bands) for $\times 4$, $\times 8$ and $\times16$ upscaling.}
	\label{tab:MS_random}
	\begin{center}
		\addtolength{\tabcolsep}{-2.7pt}
		\begin{tabular}{c || c  c | c  c | c  c | c  c | c  c | c  c | c  c | c  c}
%			\hline
			
			MS/RGB&\multicolumn{2}{c|}{Chart toy }&\multicolumn{2}{c|}{Egyptian }&\multicolumn{2}{c|}{Feathers }&\multicolumn{2}{c|}{Glass tiles }&\multicolumn{2}{c|}{Jelly beans }&\multicolumn{2}{c|}{Oil Paintings }&\multicolumn{2}{c|}{Paints }&\multicolumn{2}{c}{Average}\\ 
			\hline
			
			{$\times4$}&PSNR &SSIM & PSNR &SSIM &PSNR &SSIM &PSNR &SSIM &PSNR &SSIM &PSNR &SSIM &PSNR &SSIM &PSNR &SSIM
			\\
			\hline
%			\hline
			Bicubic &$ 28.94$&$ 0.9424$&$36.57$&$ 0.9786$&$ 30.80$&$ 0.9562$&$26.65 $&$ 0.9242$&$27.81 $&$0.9302 $&$ 31.67$&$ 0.8943$&$29.29$&$0.9493$&$30.25$&$ 0.9393$\\
			SDF~\cite{SDF} &$ 31.87$&$  0.9694$&$ 39.43$&$ 0.9795$&$ 33.45$&$ 0.9650$&$ 28.22$&$ 0.9374$&$30.32 $&$0.9433 $&$ 32.86$&$ 0.9126$&$31.96$&$0.9655$&$32.59$&$0.9532$\\
			JBF~\cite{JBU} &$ 32.56$&$  0.9653$&$ 38.73$&$ 0.9735$&$  33.60$&$ 0.9637$&$ 27.52$&$ 0.9341$&$30.29 $&$0.9498 $&$ 32.77$&$ 0.8962$&$31.94$&$0.9699$&$32.49$&$ 0.9504$\\
			JFSM~\cite{JFSM} &$ 32.98$&$ 0.9295$&$ 40.39$&$ 0.9705$&$ 33.89 $&$ 0.9425$&$ 28.98$&$ 0.9397$&$31.18 $&$ 0.9451$&$ 35.91$&$ 0.9560$&$32.76$&$0.9430$&$33.73$&$ 0.9466$\\
			GF~\cite{MSR2} &$ 34.09$&$  0.9788$&$ 40.24$&$ 0.9796$&$  33.60$&$0.9748$&$ 29.46$&$ 0.9593$&$ 30.90$&$0.9658 $&$ 35.03$&$ 0.9441$&$31.73$&$0.9702$&$33.58$&$ 0.9675$\\
			EDSR~\cite{EDSR} &$ 33.45$&$  0.9836$&$ 40.03$&$ 0.9829$&$  35.55$&$ 0.9875$&$ 29.75$&$ 0.9736$&$32.81$&$0.9838 $&$ 32.69$&$ 0.9178$&$37.28$&$0.9914$&$34.51$&$ 0.9739$\\
			SRFBN~\cite{SRFBN}  &$ 33.43$&$  0.9838$&$ 40.04$&$ 0.9822$&$  35.53$&$ 0.9873$&$ 29.53$&$ 0.9676$&$32.97$&$0.9845 $&$ 32.68$&$ 0.9182$&$36.06$&$0.9907$&$34.32$&$ 0.9735$\\
			DGF~\cite{DGF}  &$34.19$&$ 0.9559$&$37.81$&$ 0.9620$&$ 31.22 $&$0.9336 $&$29.93 $&$0.9339 $&$28.94$&$ 0.9459$&$36.10 $&$0.9649$&$ 31.72$&$ 0.9680 $&$32.84$&$0.9520$\\
%			MMSR~\cite{MMSR} &$ 37.55$&$  0.9934$&$ 45.21$&$ 0.9919$&$  39.68$&$ 0.9933$&$ 34.52$&$ 0.9912$&$38.87 $&$0.9883$&$ 37.50$&$ 0.9797$&$39.31$&$0.9943$&$38.95$&$ 0.9903$\\
			DJF~\cite{DJF} &$ 37.86$&$  0.9935$&$ 45.69$&$ 0.9922$&$  40.13$&$ 0.9939$&$ 34.97$&$ 0.9915$&$39.16 $&$0.9885$&$ 37.76$&$ 0.9805$&$39.36$&$0.9944$&$39.28$&$ 0.9906$\\
			CoISTA~\cite{JMDL} &$36.58$&$  0.9914$&$ 45.91$&$ 0.9961$&$  39.62$&$ 0.9937$&$ 33.99$&$ 0.9907$&$38.92 $&$0.9956$&$ 37.26$&$ 0.9690$&$38.40$&$0.9949$&$38.67$&$ 0.9902$\\
%			CU-Net~\cite{CUNet} &$ 39.65$&$  0.9960$&$ 47.35$&$ 0.9930$&$\textbf{42.47}$&$ 0.9965$&$\textbf{35.93}$&$\textbf{0.9944}$&$\textbf{40.69}$&$0.9972$&$38.84$&$ 0.9811$&$\textbf{40.86}$&$0.9971$&$\textbf{40.83}$&$ 0.9936$\\
%			\hline
			\hline
			LMCSC-Net&$ 40.31$&$0.9965 $&$ 48.79$&$ 0.9981$&$ 41.48$&$ 0.9962$&$ 34.65$&$ 0.9939$&$ 39.75$&$ 0.9966$&$ 39.14$&$ 0.9910$&$ 38.98$&$ 0.9966$&$ 40.44$&$ 0.9955$\\
			LMCSC$_+$-Net & $ \textbf{40.81}$&$\textbf{0.9977} $&$\textbf{49.28} $&$ 0.9989$&$ 41.53$&$ 0.9969$&$ 34.83$&$ \textbf{0.9946}$&$ \textbf{40.10}$&$ \textbf{0.9975}$&$ 38.94$&$ \textbf{0.9915}$&$ 38.92$&$ 0.9975$&$ \textbf{40.63}$&$ \textbf{0.9964}$\\
			LMCSC-ResNet &$40.47$&$0.9974$&$47.99$&$\textbf{0.9989}$&$  \textbf{41.60}$&$\textbf{0.9969}$&$ \textbf{34.96}$&$0.9943$&$39.80$&$0.9973$&$ \textbf{39.39}$&$0.9863$&$\textbf{39.53}$&$\textbf{0.9978}$&$40.53$&$0.9956$\\
			\hline
			\hline

			{$\times8$}&PSNR &SSIM & PSNR &SSIM &PSNR &SSIM &PSNR &SSIM &PSNR &SSIM &PSNR &SSIM &PSNR &SSIM &PSNR &SSIM
			\\
			\hline
%			\hline
			Bicubic &$ 25.00$&$ 0.8048$&$33.12$&$ 0.9316$&$ 25.59$&$ 0.8067$&$22.56 $&$ 0.7308$&$23.04 $&$0.7388 $&$ 30.56$&$ 0.8234$&$26.40$&$0.8465$&$26.61$&$ 0.8118$\\
			SDF~\cite{SDF} &$ 27.89$&$  0.8933$&$ 34.22$&$ 0.9366$&$ 28.25$&$ 0.8967$&$ 24.66$&$ 0.8253$&$24.88$&$0.8527 $&$ 31.61$&$ 0.8587$&$28.54$&$0.9246$&$28.58$&$0.8840$\\
			JBF~\cite{JBU} &$ 28.00$&$  0.9092$&$ 35.92$&$ 0.9578$&$  26.48$&$ 0.8492$&$ 23.06$&$ 0.7706$&$24.74 $&$0.8532 $&$ 31.90$&$ 0.8531$&$28.21$&$0.9135$&$28.33$&$ 0.8723$\\
			JFSM~\cite{JFSM} &$ 30.28$&$ 0.8897$&$ 38.31$&$ 0.9451$&$ 28.24 $&$ 0.9043$&$ 25.31$&$ 0.8851$&$26.59 $&$ 0.9083$&$ 32.55$&$ 0.9199$&$29.97$&$0.9339$&$30.18$&$ 0.9123$\\
			GF~\cite{MSR2} &$ 29.77$&$  0.9446$&$ 36.52$&$ 0.9627$&$  27.85$&$0.9183$&$ 25.24$&$ 0.8641$&$ 25.05$&$0.8813 $&$ 33.82$&$ 0.9383$&$29.25$&$0.9451$&$29.64$&$ 0.9221$\\
			EDSR~\cite{EDSR} &$ 27.04$&$  0.8802$&$ 35.62$&$ 0.9511$&$  28.38$&$ 0.9040$&$ 23.53$&$ 0.8339$&$24.73$&$0.8259 $&$ 31.95$&$ 0.8632$&$30.33$&$0.9425$&$28.78$&$ 0.8858$\\
			SRFBN~\cite{SRFBN}  &$ 27.90$&$  0.8736$&$ 35.50$&$ 0.9755$&$  30.14$&$ 0.9718$&$ 23.72$&$ 0.9002$&$25.80$&$0.9243 $&$ 31.07$&$ 0.9258$&$28.03$&$0.9653$&$28.88$&$ 0.9338$\\
			DGF~\cite{DGF}  &$28.39 $&$0.8901 $&$34.98 $&$0.9444$&$ 26.83$&$ 0.8353$&$ 25.46$&$ 0.8987$&$ 25.97$&$ 0.8852 $&$33.25 $&$0.9351$&$ 27.89$&$ 0.9128 $&$28.97$&$0.9002$\\
%			MMSR~\cite{MMSR} &$ 37.55$&$  0.9934$&$ 45.21$&$ 0.9919$&$  39.68$&$ 0.9933$&$ 34.52$&$ 0.9912$&$38.87 $&$0.9883$&$ 37.50$&$ 0.9797$&$39.31$&$0.9943$&$38.95$&$ 0.9903$\\
			DJF~\cite{DJF} &$ 32.89$&$  0.9733$&$ 41.58$&$ 0.9850$&$  31.50$&$ 0.9396$&$ 29.53$&$ 0.9685$&$30.14 $&$0.9503$&$ 35.12$&$ 0.9492$&$31.86$&$0.9553$&$33.23$&$ 0.9602$\\
			CoISTA~\cite{JMDL} &$33.18$&$  0.9768$&$ 43.46$&$ 0.9906$&$  32.04$&$ 0.9493$&$ 27.96$&$ 0.9390$&$30.69 $&$0.9585$&$ 35.99$&$ 0.9482$&$33.05$&$0.9679$&$33.77$&$ 0.9615$\\
%			CU-Net~\cite{CUNet} &$ 39.65$&$  0.9960$&$ 47.35$&$ 0.9930$&$\textbf{  42.47}$&$ 0.9965$&$\textbf{ 35.93}$&$\textbf{ 0.9944}$&$\textbf{40.69 }$&$0.9972$&$38.84$&$ 0.9811$&$\textbf{40.86}$&$0.9971$&$\textbf{40.83}$&$ 0.9936$\\
%			\hline
			\hline
			LMCSC-Net&$ 34.35$&$ 0.9805$&$ 43.90$&$ 0.9966$&$ 36.81$&$ 0.9875$&$ 30.20$&$ 0.9724$&$ 34.70$&$ 0.9888$&$ 36.27$&$ 0.9759$&$ 35.06$&$ 0.9910$&$ 35.90$&$ 0.9847$\\
            LMCSC$_+$-Net &$34.71$&$  0.9820$&$ \textbf{44.18}$&$ \textbf{0.9967}$&$  \textbf{37.36}$&$ 0.9878$&$ 30.21$&$ 0.9717$&$34.89 $&$0.9896$&$ \textbf{36.46}$&$ \textbf{0.9758}$&$\textbf{36.07}$&$\textbf{0.9930}$&$\textbf{36.27}$&$ 0.9852$\\
			LMCSC-ResNet &$\textbf{35.24}$&$\textbf{0.9835}$&$43.84$&$0.9966$&$ 37.01$&$\textbf{0.9888}$&$ \textbf{30.30}$&$\textbf{0.9756}$&$\textbf{35.07}$&$\textbf{0.9898}$&$ 36.22$&$0.9737$&$35.41$&$0.9917$&$36.16$&$\textbf{0.9857}$\\
			\hline
			\hline
			
            {$\times16$}&PSNR &SSIM & PSNR &SSIM &PSNR &SSIM &PSNR &SSIM &PSNR &SSIM &PSNR &SSIM &PSNR &SSIM &PSNR &SSIM
			\\
			\hline
%			\hline
			Bicubic &$ 22.36$&$ 0.6753$&$30.32$&$ 0.8555$&$ 23.96$&$ 0.7402$&$21.75 $&$ 0.6121$&$18.70 $&$0.4277 $&$ 27.66$&$ 0.6889$&$20.95$&$0.5975$&$23.67$&$ 0.6567$\\
			SDF~\cite{SDF} &$ 24.98$&$  0.8366$&$ 33.05$&$ 0.9298$&$ 26.87$&$ 0.8623$&$ 23.69$&$ 0.7854$&$20.89$&$0.7212 $&$ 29.69$&$ 0.8778$&$23.96$&$0.8547$&$26.16$&$ 0.8383$\\
			JBF~\cite{JBU} &$ 24.07$&$  0.8100$&$ 32.43$&$ 0.9113$&$ 25.66$&$ 0.8289$&$ 22.34$&$ 0.6996$&$19.97 $&$0.6264 $&$ 28.87$&$ 0.7747$&$23.14$&$0.7894$&$ 25.21$&$  0.7772$\\
			JFSM~\cite{JFSM} &$ 27.54$&$ 0.8617$&$ 36.01$&$ 0.9294$&$ 27.79 $&$ 0.8750$&$ 25.04$&$ 0.8682$&$24.05 $&$ 0.8773$&$ 31.76$&$ 0.9104$&$27.38$&$0.8400$&$ 28.51$&$  0.8803$\\
			GF~\cite{MSR2} &$25.80$& $0.8834$& $33.71 $& $0.9347$& $ 27.58$& $ 0.8840$& $ 24.54$& $ 0.8130$& $ 21.41$& $ 0.7912$& $ 31.87$& $ 0.9096$& $ 25.54$& $ 0.8822$&$27.21$&$0.8712$\\
%			EDSR~\cite{EDSR} &$ 27.04$&$  0.8802$&$ 35.62$&$ 0.9511$&$  28.38$&$ 0.9040$&$ 23.53$&$ 0.8339$&$24.73$&$0.8259 $&$ 31.95$&$ 0.8632$&$30.33$&$0.9425$&$-$&$ -$\\
			SRFBN~\cite{SRFBN}  &$ 23.78$&$  0.8203$&$ 32.32$&$ 0.9452$&$  25.11$&$ 0.8816$&$ 22.49$&$ 0.7643$&$20.70$&$0.7144 $&$ 28.78$&$ 0.8483$&$24.78$&$0.9085$&$25.42$&$ 0.8404$\\
			DGF~\cite{DGF}  &$26.83 $&$0.8330 $&$33.64$&$ 0.9378 $&$25.25 $&$0.8127 $&$24.16 $&$0.8690$&$ 23.23$&$ 0.8543$&$ 31.84 $&$0.9332$&$ 26.16$&$ 0.8654 $&$27.30$&$0.8722$\\
%			MMSR~\cite{MMSR} &$ 37.55$&$  0.9934$&$ 45.21$&$ 0.9919$&$  39.68$&$ 0.9933$&$ 34.52$&$ 0.9912$&$38.87 $&$0.9883$&$ 37.50$&$ 0.9797$&$39.31$&$0.9943$&$38.95$&$ 0.9903$\\
			DJF~\cite{DJF} &$ 30.03$&$ 0.9582 $&$38.82$&$ 0.9692$&$31.03 $&$0.9370 $&$\textbf{29.08}$&$ \textbf{0.9660} $&$25.43 $&$0.8992 $&$33.04 $&$0.9171 $&$30.58 $&$0.9401$&$28.59$&$ 0.9174$\\
			CoISTA~\cite{JMDL} &$29.85$&$ 0.9500$&$ 39.18$&$ 0.9760$&$ 31.07$&$ 0.9433 $&$27.82 $&$0.9231$&$26.66$&$ 0.9013$&$ \textbf{33.95}$&$ 0.9239$&$ 31.06 $&$0.9484
            $&$29.50$&$0.9258$\\
%			CU-Net~\cite{CUNet} &$ 39.65$&$  0.9960$&$ 47.35$&$ 0.9930$&$\textbf{  42.47}$&$ 0.9965$&$\textbf{ 35.93}$&$\textbf{ 0.9944}$&$\textbf{40.69 }$&$0.9972$&$38.84$&$ 0.9811$&$\textbf{40.86}$&$0.9971$&$\textbf{40.83}$&$ 0.9936$\\
%			\hline
			\hline
			LMCSC-Net &$ 31.52$&$ 0.9716$&$ 41.43$&$ 0.9916$&$ 32.29$&$ 0.9655$&$ 27.29$&$ 0.9328$&$28.42$&$ 0.9632$&$ 32.96$&$ \textbf{0.9421}$&$ 31.14$&$ 0.9757$&$ 32.14$&$ 0.9632$\\
            {LMCSC$_+$-Net} & $31.31$ & $0.9688$ & $41.06$ &$\textbf{0.9928}$ &$32.45$ &$\textbf{0.9694}$ &$27.16$ &$0.9310$ &$28.72$ &$\textbf{0.9640}$ &$32.23$ &$0.9364$ &$\textbf{31.25}$ &$\textbf{0.9784}$ &$32.03$ &$0.9630$ \\
			LMCSC-ResNet &$\textbf{31.69}$&$\textbf{0.9721}$&$\textbf{41.51}$&$0.9923$&$  \textbf{32.67}$&$0.9692$&$ 27.48$&$0.9336$&$\textbf{28.75}$&$0.9624$&$ 33.16$&$0.9413$&$30.44$&$0.9746$&$\textbf{32.24}$&$\textbf{0.9636}$\\
			\hline
			\hline
		\end{tabular}
	\end{center}
\end{table*}

%\begin{figure*}[t!]
%	\centering
%	\includegraphics[width=1\textwidth]{figures/hist-03-50.png}\\
%	\caption{Histograms of the sparse feature maps $U$ and $U-Z$ for two NIR images u-0040 and u-0050. The visualizations clearly show that the network is able to follow the specific structure of the problem where $U$ and $U-Z$ are supposed to be drawn from the Laplace distribution. }
%	\label{fig:hist}
%\end{figure*}

\begin{figure*}
	\centering
	\subfloat[]{\includegraphics[scale=0.21]{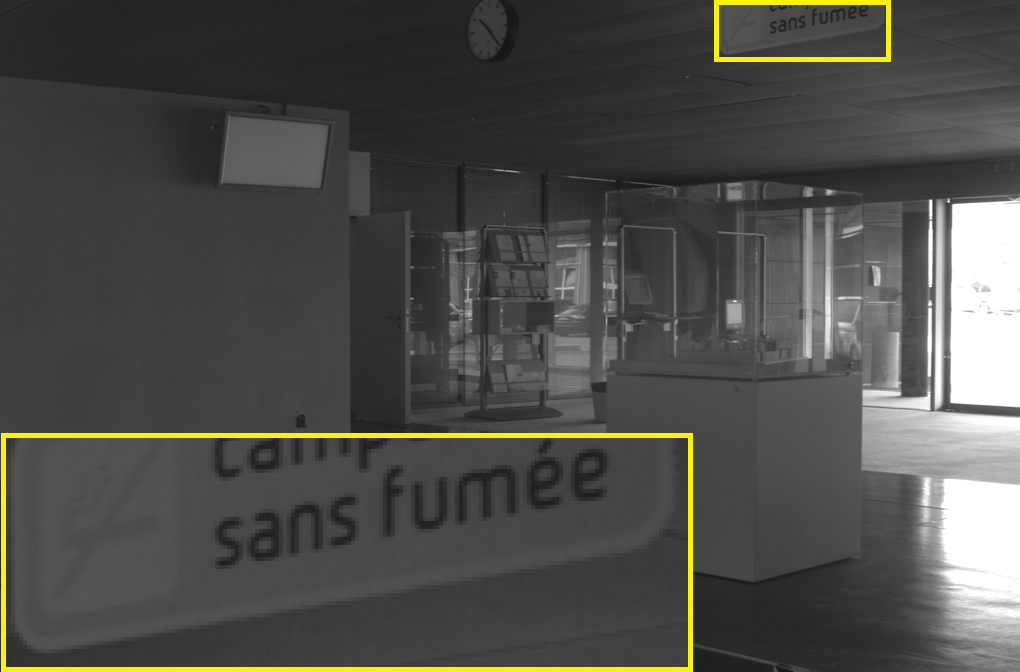}}
	\hspace{0.2cm}
	\subfloat[]{\includegraphics[scale=0.21]{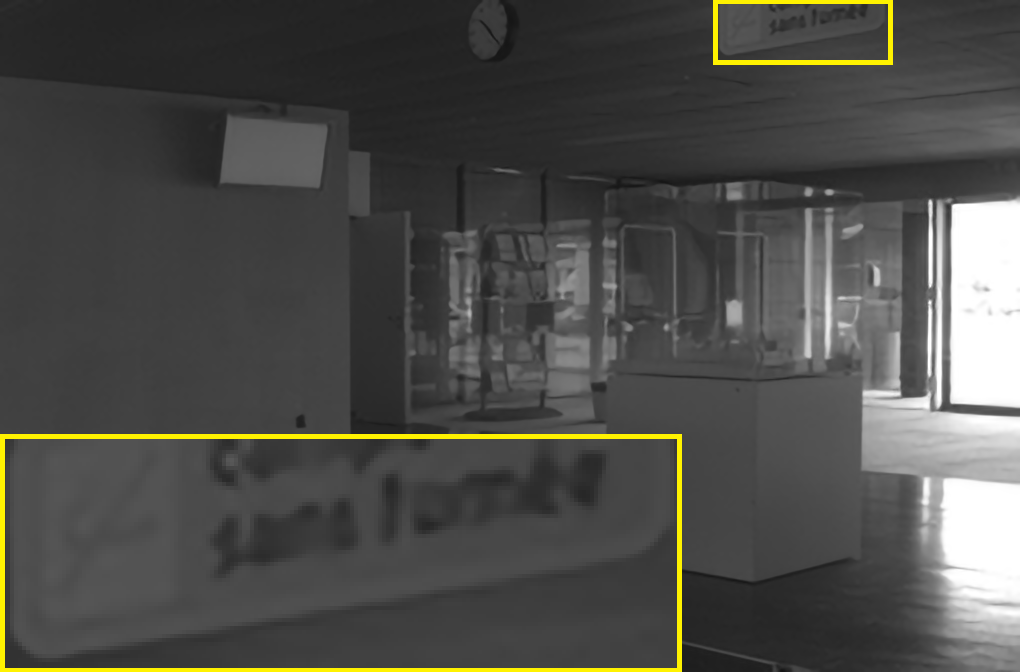}}
	\hspace{0.2cm}
	\subfloat[]{\includegraphics[scale=0.21]{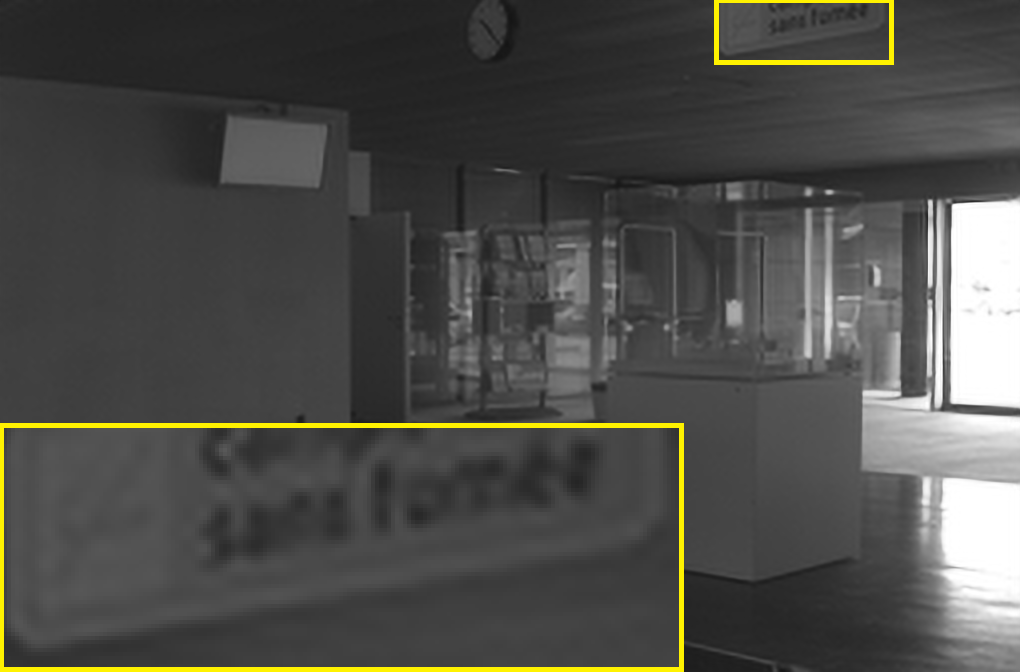}}
	\hspace{0.2cm}
	\subfloat[]{\includegraphics[scale=0.21]{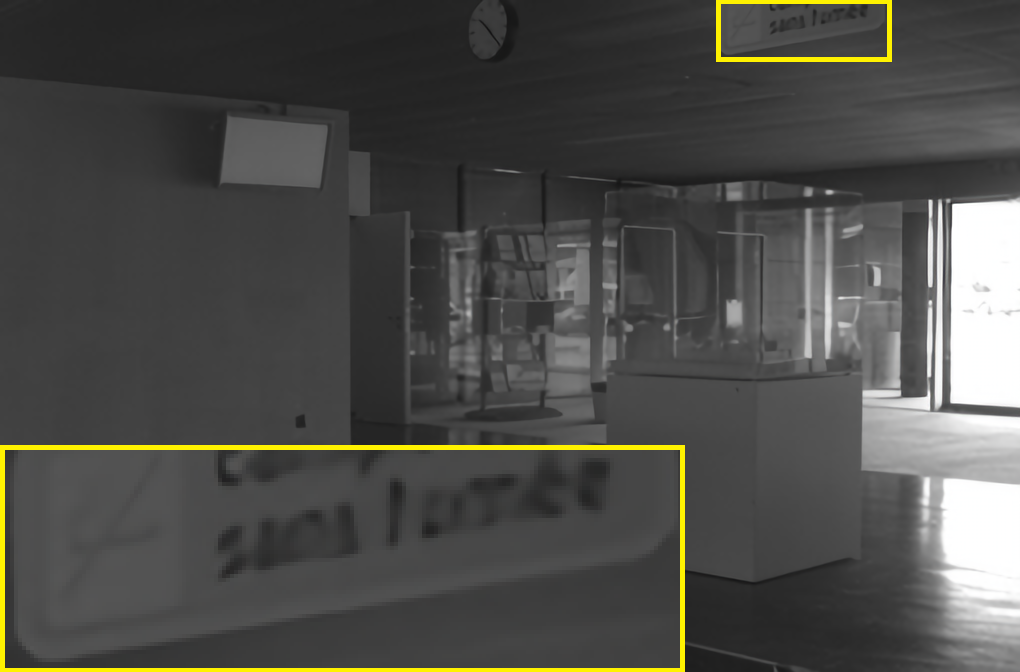}}
	\hspace{0.2cm}
	\subfloat[]{\includegraphics[scale=0.21]{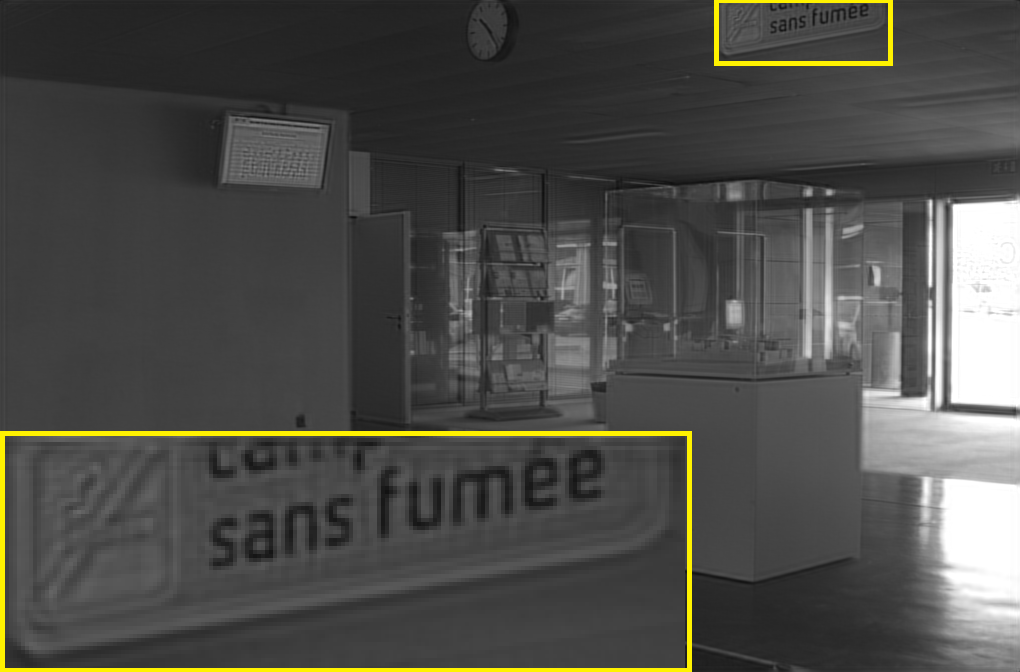}}
	\hspace{0.2cm}
	\subfloat[]{\includegraphics[scale=0.21]{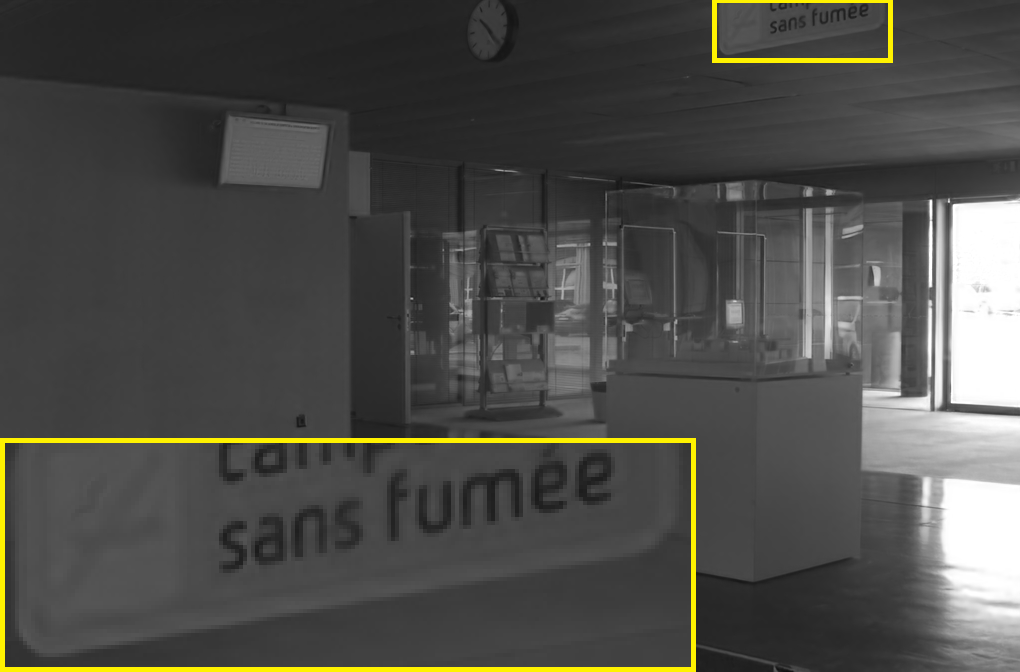}}
	\caption{Super-resolution of the NIR image ``i-0013" with upscaling factor $\times4$. (a)~ground-truth (b)~CSCN~\cite{Huang}, (c)~ACSC~\cite{raja}, (d)~EDSR~\cite{EDSR}, (e)~DJF~\cite{DJF} and (f)~LMCSC-ResNet. }   
	\label{fig:visualEx2}
\end{figure*}

\subsubsection{Super-resolution of NIR/RGB images}
The first set of experiments involves super-resolution of NIR images with the aid of RGB images.
The proposed models are compared with several single-modal
and multimodal methods. 
Among the reference single-modal methods, the network designs proposed in~\cite{Huang} and \cite{raja}
also involve deep unfolding architectures with sparse priors;\footnote{The model 
in~\cite{raja} has been applied to image denoising and inpainting;
thus, we slightly modified it and trained it for SR. 
The number and size of convolutional filters 
is different than what reported in~\cite{raja} and 
we do not force the layers to share weights.} 
\cite{EDSR}~is a residual network, while \cite{SRFBN}~has an RNN structure.
The multimodal designs include~\cite{DJF} and~\cite{JMDL};
recall that the latter is a LISTA-based network.

\begin{figure*}
	\centering
	\subfloat[]{\includegraphics[scale=0.37]{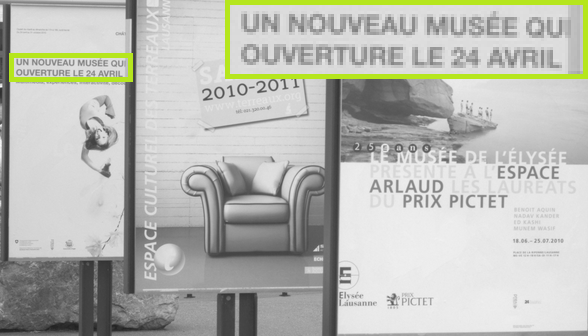}}
	\hspace{0.2cm}
	\subfloat[]{\includegraphics[scale=0.37]{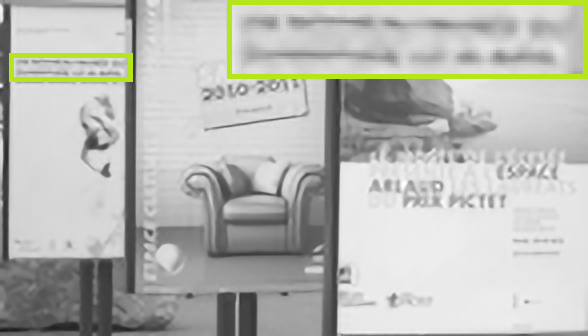}}
	\hspace{0.2cm}
	\subfloat[]{\includegraphics[scale=0.37]{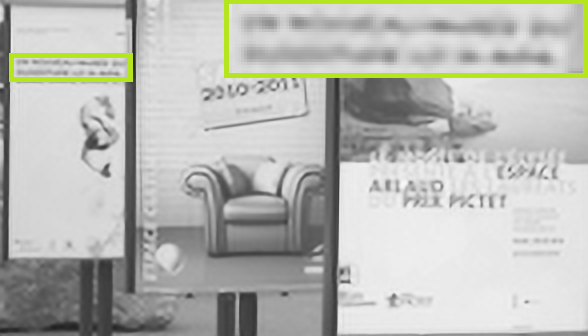}}
	\hspace{0.2cm}
	\subfloat[]{\includegraphics[scale=0.37]{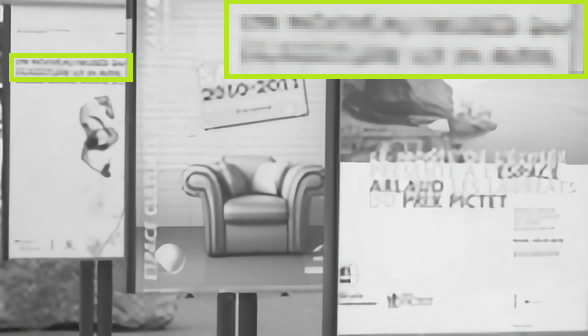}}
	\hspace{0.2cm}
	\subfloat[]{\includegraphics[scale=0.37]{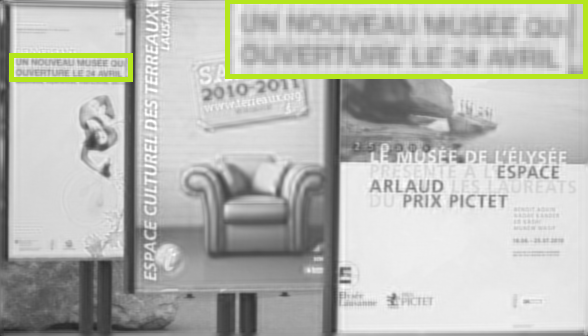}}
	\hspace{0.2cm}
	\subfloat[]{\includegraphics[scale=0.37]{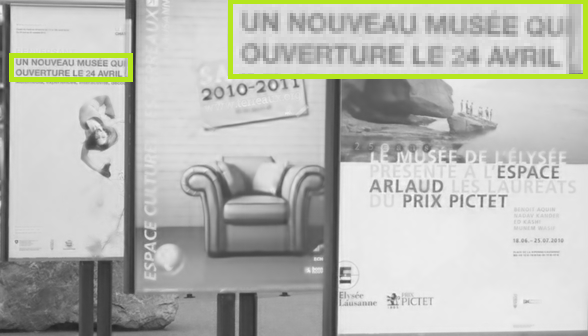}}
	\caption{Super-resolution of the test NIR image ``u-0000" with upscaling factor $\times4$. (a)~ground-truth (b)~CSCN~\cite{Huang}, (c)~ACSC~\cite{raja}, (d)~EDSR~\cite{EDSR}, (e)~DJF~\cite{DJF} and (f)~LMCSC-ResNet. Results for CDLSR~\cite{MSR} are not presented as the code is not available.}  
	\label{fig:visualEx1}
\end{figure*}

We employ the EPFL RGB-NIR dataset and test our models on $25$ NIR/RGB image pairs.
Table~\ref{tab:NIR_25} presents average results  in terms of the Peak Signal-to-Noise Ratio (PSNR)
for upscaling factors equal to $\times 2$, $\times 4$ and $\times 6$. 
We also present detailed PSNR and structural similarity (SSIM) index results for selected test images\footnote{A test image
is identified by a letter \textquotedblleft{u}\textquotedblright, 
\textquotedblleft{o}\textquotedblright,  \textquotedblleft{i}\textquotedblright \ referring to 
the folders \texttt{urban},  \texttt{oldbuilding}, \texttt{indoor} in the dataset, 
and a code \textquotedblleft{00xx}\textquotedblright.} 
in~Table~\ref{tab:NIR_8};
these images have been used for testing in~\cite{MSR}.
As can be seen in Table~\ref{tab:NIR_25}, 
our models deliver higher reconstruction accuracy compared to reference methods at all scales, 
with LMCSC-ResNet achieving the best performance.
Furthermore, the numerical results show that as the upscaling factor increases, 
the PSNR gain of LMCSC-ResNet over the second best reference method~\cite{JMDL} also increases. 
For instance, at scale $\times2$ the gain is $0.64$dB, and rises to $0.84$dB and $1.45$dB 
for scales $\times4$ and $\times6$, respectively. 
The results for selected images in Table~\ref{tab:NIR_8} are similar.
The average PSNR gain over~\cite{JMDL} is $1.7$dB at scale $\times2$,
and grows to $1.7$dB  at scale $\times4$, and $2.03$dB at scale $\times6$. 
The role of multimodal fusion becomes more significant as the SR upscaling factor increases; 
our numerical analysis shows that the proposed models 
can effectively fuse the information from two different modalities.

\subsubsection{Depth upsampling}
For the application of depth map upsampling with the aid of RGB images, 
we train our LMCSC-ResNet network 
for three upscaling factors, $\times4$, $\times8$ and $\times16$,
using the first $1000$ images of the NYU v2 dataset~\cite{NYU}. 
We report averaged Root Mean Square Error (RMSE) results 
over $449$ test images in Table~\ref{tab:NYU_449} 
and also detailed results for $10$ selected images from this dataset in Table~\ref{tab:NYU_10}.
Table~\ref{tab:NYU_449} involves comparison with several multimodal methods such as 
the joint filtering approaches JBF~\cite{JBU}, GF~\cite{MSR2},  SDF~\cite{SDF},
a learning based method proposed by Gu \textit{et al.}~\cite{dynamic-guidance},
and the deep learning design DJF~\cite{DJF}; 
the numerical results for the reference methods are provided by the authors of~\cite{dynamic-guidance}. 
We select the second and third best methods from Table~\ref{tab:NYU_449},
that is,~\cite{DJF} and \cite{dynamic-guidance},
to compare the performance of our models on selected images in Table~\ref{tab:NYU_10}.
We note that for the scales $\times4$ and $\times8$,  
LMCSC$_+$-Net achieves the lowest average RMSE. 
However, for an upscaling factor $\times16$ 
LMCSC-ResNet has the best performance. 
At the highest upscaling factor, in the presence of very limited information from the LR input, 
LMCSC-ResNet has an RMSE gain of $0.36$ over the second best method, i.e., Gu \textit{et al.}~\cite{dynamic-guidance}. 

\subsubsection{Multi-spectral data super-resolution}
We utilize the Columbia multi-spectral dataset for the last set of experiments. 
We apply LMCSC-ResNet to super-resolve spectral images using their RGB version as side information.
The experiments involve comparison of our models
against  several multimodal methods,
namely, JBF~\cite{JBU}, GF~\cite{MSR2} and SDF~\cite{SDF}, 
which are joint filtering approaches,
DJF~\cite{DJF} and CoISTA~\cite{JMDL}
which are deep learning designs,
the method proposed in~\cite{JFSM} coined JFSM,  
and the coupled dictionary learning based method~CDLSR~\cite{MSR}.
We also report results for the single-modal deep learning designs 
EDSR~\cite{EDSR} and SRFBN~\cite{SRFBN}.
Numerical results  are presented in Tables~\ref{tab:MS_640nm} and~\ref{tab:MS_random}.
Table~\ref{tab:MS_640nm} includes results for $\times4$ upsampling in terms of PSNR and SSIM 
for $7$ test images of the $640$nm band. 
We can see that LMCSC-ResNet yields the best performance for scale $\times4$ with a gain of more than $2.1$dB over~\cite{JMDL}. 
Table~\ref{tab:MS_random} presents detailed results for three different scales, $\times4$, $\times8$ and $\times16$, 
for randomly selected test images from different bands. 
Similar to the depth map upsampling application, 
Table~\ref{tab:MS_random} shows that for $\times4$ and $\times8$ scales  
LMCSC$_+$-Net provides the highest average PSNR. 
For $\times16$, LMCSC-ResNet delivers the best reconstruction accuracy 
with a PSNR gain of $2.74$dB over~\cite{JMDL}, the second best method among the competing works.

\begin{figure*}
	\centering
	\subfloat[]{\includegraphics[scale=0.26]{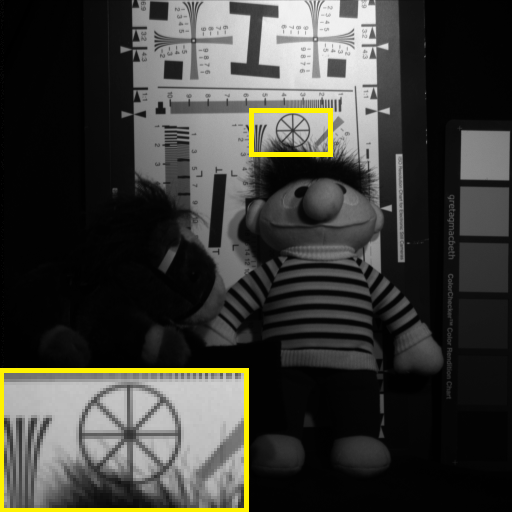}}
	%	\hspace{0.00cm}
	%	\subfloat[]{\includegraphics[scale=0.224]{figures/chart_BC_X4.png}}
	\hspace{0.00cm}
	\subfloat[]{\includegraphics[scale=0.26]{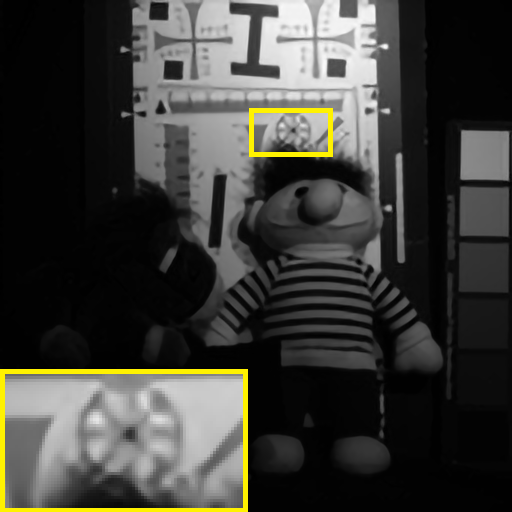}}
	\hspace{0.00cm}
	\subfloat[]{\includegraphics[scale=0.26]{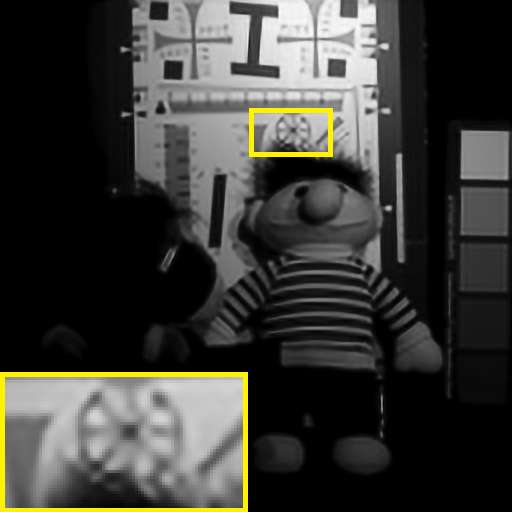}}
	\hspace{0.00cm}
	\subfloat[]{\includegraphics[scale=0.26]{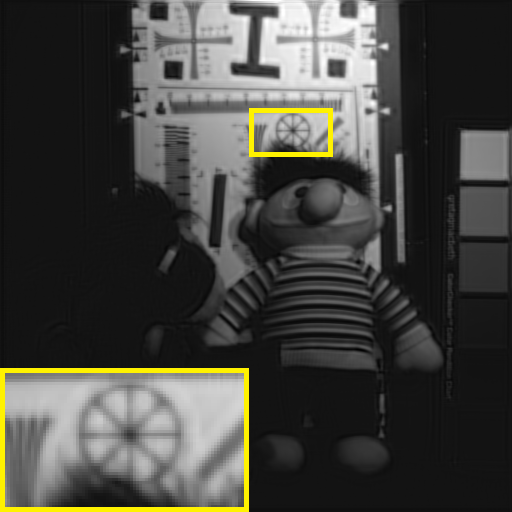}}
	\hspace{0.00cm}
	\subfloat[]{\includegraphics[scale=0.26]{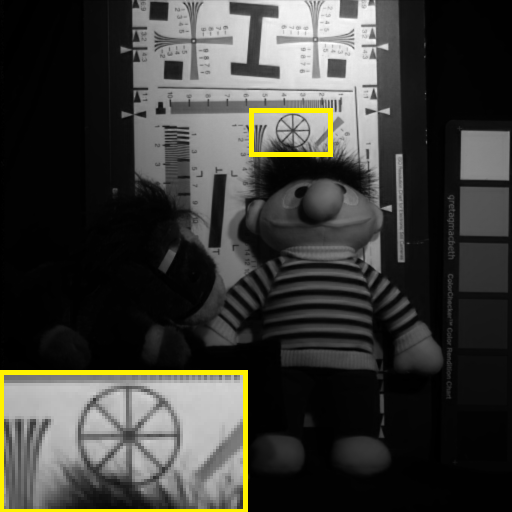}}
	
	\caption{Super-resolution of the test multi-spectral image ``chart" with upscaling factor $\times4$. (a)~ground-truth, (b)~CSCN~\cite{Huang}, (c)~ACSC~\cite{raja}, (d)~DJF~\cite{DJF} and (e)~LMCSC-ResNet.}   
	\label{fig:visualEx3}
\end{figure*}

\subsubsection{Visual examples}
\begin{figure*}
	\centering
	\subfloat[]{\includegraphics[scale=0.26]{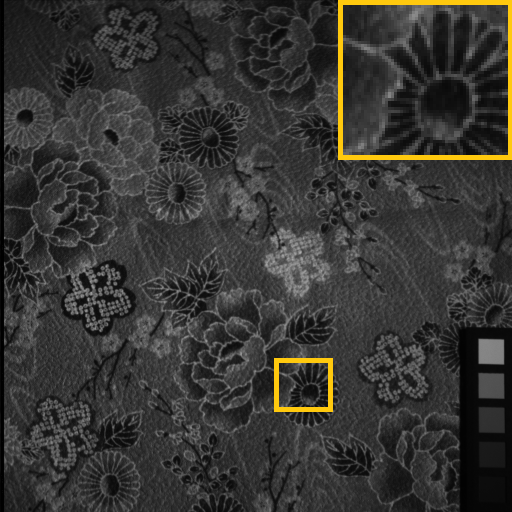}}
	\hspace{0.01cm}
	%	\subfloat[]{\includegraphics[scale=0.21]{figures/cloth_BC_X4.png}}
	%	\hspace{0.01cm}
	\subfloat[]{\includegraphics[scale=0.26]{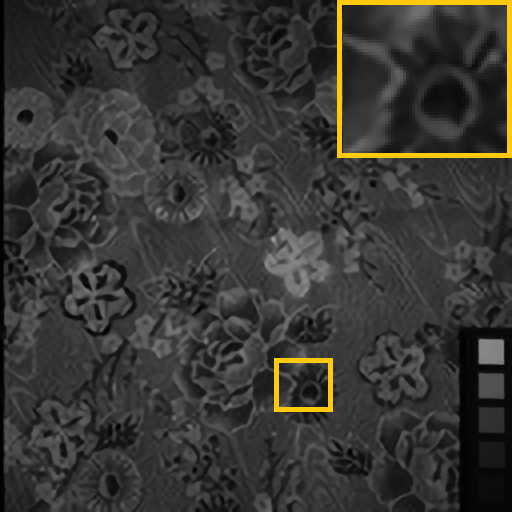}}
	\hspace{0.01cm}
	\subfloat[]{\includegraphics[scale=0.26]{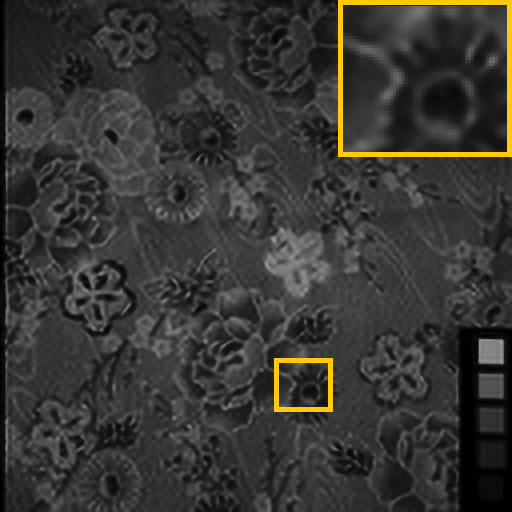}}
	\hspace{0.01cm}
	\subfloat[]{\includegraphics[scale=0.26]{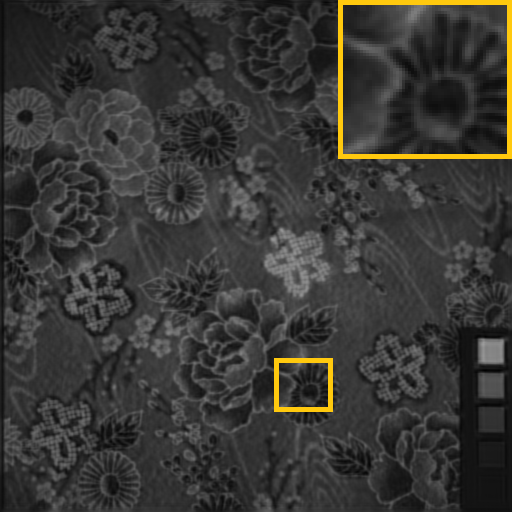}}
	\hspace{0.01cm}
	\subfloat[]{\includegraphics[scale=0.26]{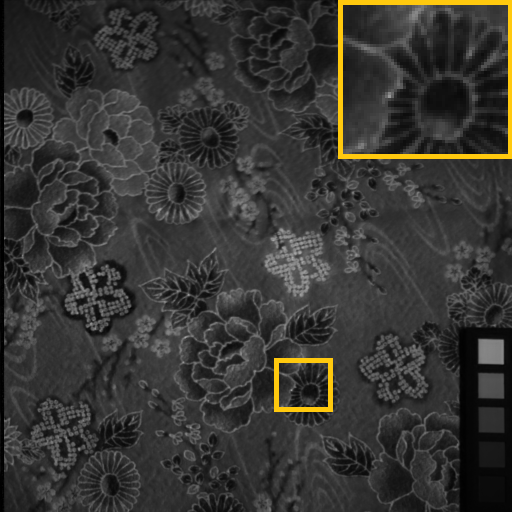}}
	
	\caption{Super-resolution of the test multi-spectral image ``cloth" with upscaling factor $\times4$. (a)~ground-truth, (b)~CSCN~\cite{Huang}, (c)~ACSC~\cite{raja}, (d)~DJF~\cite{DJF} and (e)~LMCSC-ResNet.}   
	\label{fig:visualEx4}
\end{figure*}

We conclude our experiments with visual examples for two super-resolved NIR images, 
presented in Fig.~\ref{fig:visualEx1} and Fig.~\ref{fig:visualEx2},
as well as for two  super-resolved multi-spectral images, 
presented in Fig.~\ref{fig:visualEx3} and Fig.~\ref{fig:visualEx4}.
The visual comparison corroborates the presented numerical results.

\begin{table}[t!]
	\centering
	\caption{Comparison of the inference time (sec) of our multimodal architectures and CoISTA~\cite{JMDL} w.r.t. two input resolutions. }
	\label{tab:complexity}
	\begin{center}
		\addtolength{\tabcolsep}{-2.4pt}
		\begin{tabular}{c  | c  c   c    c}
%			\hline
			
			input size& CoISTA~\cite{JMDL}&\multicolumn{1}{c}{LMCSC-Net}&\multicolumn{1}{c}{LMCSC$_+$-Net}&\multicolumn{1}{c}{LMCSC-ResNet}\\ 
			\hline

			{256$\times2$56}&$0.51$&$1.09$ &$ 1.14$&$ 1.14$\\

%			\hline
%			\hline
			{512$\times$512}&$0.63$&$1.29$ &$ 1.38$&$ 1.39$\\

			\hline

		\end{tabular}
	\end{center}
\end{table}

\subsection{Complexity Analysis} 
The proposed LMCSC-based networks operate on the entire image  rather than reconstructing overlapping patches and aggregating them. 
The inference time of the proposed networks is independent of the upscaling factor, as the input is first upscaled to the desired resolution using bicubic interpolation. 
Table~\ref{tab:complexity} reports averaged inference running times of the proposed models and CoISTA~\cite{JMDL} on $10$ test images from the Columbia multi-spectral database with size $512\times512$ pixels and on $10$ images of size $256\times256$ pixels, which are cropped versions of the previous ones. 
All models are tested on an NVIDIA GeForce GTX 1070. 
We observe that our proposed networks have higher inference times compared to CoISTA~\cite{JMDL}, which is due to the proposed fusion strategy. Specifically, CoISTA~\cite{JMDL} fuses modalities in the last layer of the network by a linear combination of the sparse representations. In contrast, the LMCSC layers of our networks perform intermediate fusion using the activation function defined in~\eqref{eq:prox1} and~\eqref{eq:prox2}. As demonstrated by the reconstruction accuracy results, where gains of up to $2.74$dB over CoISTA~\cite{JMDL} were reported, the proposed fusion approach can effectively capture the complex relationships across modalities at the expense of a reasonable increase in terms of computation.

%%%%%%%%%%%%%%%%%%%%%%%%%%%%%%%%%%
\section{Conclusions}
\label{sec:conclusion}
%%%%%%%%%%%%%%%%%%%%%%%%%%%%%%%%%%
In this paper, we presented a new approach for guided image super-resolution
based on a novel multimodal deep unfolding design.
The efficient integration of the guidance modality into the deep learning architecture is achieved
with a neural network that performs steps similar to
an iterative algorithm for convolutional sparse coding with side information.
By exploiting residual learning,
we further improve the training efficiency and increase reconstruction accuracy of the proposed framework.
Our approach was applied to super-resolution of NIR images, multi-spectral and depth maps with the aid of HR RGB images.
The superior performance of our models against various state-of-the-art single-modal and multimodal methods
was demonstrated by experimental results.



\bibliographystyle{IEEEtran}
\bibliography{IEEEabrv,refs}
\end{document}